\documentclass[a4paper,12pt]{article}
\pdfoutput=1
\usepackage{graphicx, rotating}
\usepackage{hyperref}
\usepackage{slashed}
\usepackage{ifpdf}

\ifx\pdfoutput\undefined
\usepackage[dvips,bookmarks=false]{hyperref}	% This is for arXiv.org
\else
\usepackage{hyperref}	% This is for pdftex
\fi
\hypersetup{colorlinks,bookmarksopen,bookmarksnumbered,citecolor=verdes,
linkcolor=blus,pdfstartview=FitH,urlcolor=rossos}
\def\hhref#1{\href{http://arxiv.org/abs/#1}{#1}} % in bibliography
\usepackage{amsfonts}
\usepackage{amsmath}
\usepackage{slashed}

\newcommand{\beq}{\begin{equation}}
\newcommand{\eeq}{\end{equation}}
\newcommand{\fig}[1]{~\ref{fig:#1}}

\newcount\Mac  \Mac=1  % devo mettere Mac=1 se sto lavorando sul file Mac
\newcommand{\ifMac}[2]{\ifnum\Mac=1 #1 \else #2 \fi}
\def\putps(#1,#2)(#3,#4)#5#6{\ifnum\Mac=1 \put(#1,#2){\special{picture #5}}
\else  \put(#3,#4){\includegraphics{#6}} \fi}

\newcommand{\One}{\hbox{1\kern-.24em I}}

\newcommand{\GeV}{\,{\rm GeV}}
\newcommand{\TeV}{\,{\rm TeV}}

\newcommand{\eq}[1]{~{\rm (\ref{eq:#1})}}

\newcommand{\lascia}[1]{}
\makeatletter
\def\art{\@ifnextchar[{\eart}{\oart}}
\def\eart[#1]#2#3#4#5#6{{\rm #2}, {#3 #4} {\rm (#6) #5} [arXiv:{\hhref{#1}}]}
\def\hepart[#1]#2{{\rm #2, arXiv:\hhref{#1}}}
\newcommand{\oart}[5]{{\rm #1}, {#2 #3} {\rm (#5) #4}}

\newcounter{alphaequation}[equation]
\def\thealphaequation{\theequation\hbox to
0.6em{\hfil\alph{alphaequation}\hfil}}
\def\eqnsystem#1{
\def\@eqnnum{{\rm (\thealphaequation)}}
\def\@@eqncr{\let\@tempa\relax \ifcase\@eqcnt \def\@tempa{& & &} \or
  \def\@tempa{& &}\or \def\@tempa{&}\fi\@tempa
  \if@eqnsw\@eqnnum\refstepcounter{alphaequation}\fi
\global\@eqnswtrue\global\@eqcnt=0\cr}
\refstepcounter{equation} \let\@currentlabel\theequation \def\@tempb{#1}
\ifx\@tempb\empty\else\label{#1}\fi
\refstepcounter{alphaequation}
\let\@currentlabel\thealphaequation
\global\@eqnswtrue\global\@eqcnt=0 \tabskip\@centering\let\\=\@eqncr
$$\halign to \displaywidth\bgroup \@eqnsel\hskip\@centering
$\displaystyle\tabskip\z@{##}$&\global\@eqcnt\@ne
\hskip2\arraycolsep\hfil${##}$\hfil& \global\@eqcnt\tw@\hskip2\arraycolsep
$\displaystyle\tabskip\z@{##}$\hfil
\tabskip\@centering&\llap{##}\tabskip\z@\cr}

\def\endeqnsystem{\@@eqncr\egroup$$\global\@ignoretrue} \makeatother

\def\Lag{{\cal L}}

\def\circa#1{\,\raise.3ex\hbox{$#1$\kern-.75em\lower1ex\hbox{$\sim$}}\,}

\usepackage{multicol}
\usepackage{color}
\definecolor{rosso}{cmyk}{0,1,1,0.4}
\definecolor{rossos}{cmyk}{0,1,1,0.55}
\definecolor{rossoc}{cmyk}{0,1,1,0.2}
\definecolor{blu}{cmyk}{1,1,0,0.3}
\definecolor{blus}{cmyk}{1,1,0,0.6}
\definecolor{bluc}{cmyk}{1,1,0,0.1}
\definecolor{verde}{cmyk}{0.92,0,0.59,0.25}
\definecolor{verdec}{cmyk}{0.92,0,0.59,0.15}
\definecolor{verdes}{cmyk}{0.92,0,0.59,0.4}
\definecolor{grigio}{cmyk}{0,0,0,0.07}
\definecolor{rosa}{cmyk}{0,0.1,0.1,0.02}
\definecolor{rosino}{cmyk}{0,0.05,0.05,0.02}
\definecolor{rosas}{cmyk}{0,0.3,0.25,0.05}
\definecolor{celeste}{cmyk}{0.1,0,0,0.02}
\definecolor{giallino}{cmyk}{0,0,0.4,0.02}
\definecolor{rosso}{cmyk}{0,1,1,0.4}
\definecolor{rossos}{cmyk}{0,1,1,0.55}
\definecolor{rossoc}{cmyk}{0,1,1,0.2}
\definecolor{blu}{cmyk}{1,1,0,0.3}
\definecolor{bluc}{cmyk}{1,1,0,0.1}
\definecolor{blucc}{cmyk}{0.7,0.5,0,0}
\definecolor{viola}{cmyk}{0,1,0,0.6}
\definecolor{viola2}{cmyk}{0,1,0.2,0.6}
\definecolor{verde}{cmyk}{0.92,0,0.59,0.25}
\definecolor{verdec}{cmyk}{0.92,0,0.59,0.15}
\definecolor{verdes}{cmyk}{0.92,0,0.59,0.4}
\definecolor{verdino}{cmyk}{0.12,0,0.09,0.05}
\definecolor{giallo}{cmyk}{0,0,1,0}
\definecolor{gialloverde}{cmyk}{0.44,0,0.74,0}

\font\tenrsfs=rsfs10 at 12pt

\font\sevenrsfs=rsfs7
\font\fiversfs=rsfs5
\newfam\rsfsfam
\textfont\rsfsfam=\tenrsfs
\scriptfont\rsfsfam=\sevenrsfs
\scriptscriptfont\rsfsfam=\fiversfs
\def\mathscr#1{{\fam\rsfsfam\relax#1}}
\def\Lag{\mathscr{L}}

\begin{document}
% IFUP-TH/2011-1\hfill CERN-PH-TH/2010-XX
\color{black}
\vspace{0.5cm}
\begin{center}
{\Huge\bf\color{rossos}Implications of {\sc Xenon}100 and LHC\\[3mm]  results  for Dark Matter models\\[5mm]
\LARGE (updated including 2012 data)}\\
\bigskip\color{black}\vspace{0.6cm}
{{\large\bf Marco Farina$^{a}$,  Mario Kadastik$^{b}$,  Duccio Pappadopulo$^{c}$,\\[2mm]
 Joosep Pata$^{d}$, Martti Raidal$^{b}$, Alessandro Strumia$^{b,e}$}
} \\[7mm]
{\it  (a)  Scuola Normale Superiore and INFN, Piazza dei Cavalieri 7, 56126 Pisa, Italia}\\[3mm]
{\it  (b) National Institute of Chemical Physics and Biophysics, Ravala 10, Tallinn, Estonia}\\[3mm]
{\it  (c) Institut de Th\'eorie des Ph\'enom\`enes Physiques, EPFL,  CH--1015 Lausanne, Switzerland}\\[3mm]
{\it  (d) Department of Physics, University of Tartu, Estonia}\\[3mm]
{\it  (e) Dipartimento di Fisica dell'Universit{\`a} di Pisa and INFN, Italia}\\[3mm]
%{\it CERN, PH-TH, CH-1211, Geneva 23, Switzerland}\\[3mm]
\end{center}
\bigskip
\centerline{\large\bf\color{blus} Abstract}
\begin{quote}
We perform a fit to the recent {\sc Xenon}100 data and study its implications for Dark Matter scenarios.
We find that Inelastic Dark Matter is disfavoured as an explanation to the DAMA/LIBRA annual modulation signal. 
Concerning the scalar singlet DM model, we find that the {\sc Xenon}100 data disfavors its constrained limit.
We study the CMSSM as well as the low scale phenomenological 
MSSM taking into account latest Tevatron and LHC data (1.1/fb) about sparticles and $B_s\to\mu\mu$.
After the EPS 2011 conference, LHC excludes the ``Higgs-resonance'' region of DM freeze-out 
and {\sc Xenon}100 disfavors the ``well-tempered" bino/higgsino,  realized in the ``focus-point" region of the CMSSM parameter space.
The preferred region shifts to heavier sparticles, higher fine-tuning, higher $\tan\beta$ and the quality of the fit deteriorates.
%The implications of the complementary LHC and {\sc Xenon}100 data for Dark Matter annihilation scenarios is outlined. 
\color{black}
\end{quote}

\section{Introduction}

The best motivated candidates for the cold Dark Matter (DM) of the Universe~\cite{WMAP7}  are Weakly Interacting Massive Particles (WIMPs)~\cite{WIMP}.
Indeed, a stable particle with a typical weak interaction cross section of 1~pb and a mass of order 100~GeV  naturally
produces the observed DM thermal relic abundance.   Because the evidence for DM is 
purely gravitational, there are many particle physics models for WIMPs. Theoretically most appealing ones among those are 
the DM models based on supersymmetry (SUSY)~\cite{EllisDM} in which the DM is stable due to the existence of discrete $Z_2$ symmetry -- 
the R-parity~\cite{RP} or equivalently the matter parity~\cite{MP}. 
Because the matter parity may occur also in the scalar extensions of the standard model (SM)~\cite{Kadastik:2009dj},  
the scalar DM models are also well motivated from the underlying unified physics point of view.  

The WIMPs have been searched for in the LEP, Tevatron and LHC experiments as well as in the underground DM direct detection experiments 
that detect  DM recoils on nuclei. The history of DM direct detection experiments can be characterized with their sensitivity to the 
spin-independent DM-nucleon interaction cross section $\sigma_{\rm SI}$ as follows.
\begin{enumerate}
\item The  non-relativistic DM annihilation cross section suggested by the observed cosmological DM abundance is
 $\sigma \sim 1/(T_0M_{\rm Pl}) \approx 3~10^{-26}\,{\rm cm}^3/{\rm s}=
10^{-36}\,{\rm cm}^2$. Such a large value of  $\sigma_{\rm SI}$ is clearly excluded unless the DM particles have peculiar properties
or their interactions are restricted to untested sectors.
\item DM that interacts with the $Z$ boson: $\sigma_{\rm SI} \approx \alpha^2 m_N^2/M_Z^4\approx 10^{-38}\,{\rm cm}^2$. Again, 
generic WIMPs of this type have been excluded.
\item DM that interacts with the Higgs boson: $\sigma_{\rm SI} \approx \alpha^2 m_N^4/M_{\rm DM}^2M_Z^4\approx 10^{-43}\,{\rm cm}^2$, 
having assumed that the DM-DM-Higgs coupling is comparable with the weak coupling $\alpha$.

\item DM that interacts with the $W^\pm$ boson and consequently at loop level with nucleons~\cite{Hisano:2011cs}, 
$\sigma_{\rm SI} \approx \alpha^4 m_N^4/(4\pi)^2M_{\rm DM}^2M_W^6\approx 10^{-46}\,{\rm cm}^2$, as predicted e.g.\ by Minimal Dark Matter~\cite{MDM}.
\end{enumerate}
Therefore, in the absence of new gauge bosons, 
the present most sensitive DM direct search experiments like CDMS~\cite{CDMS}, EDELWEISS~\cite{EDELWEISS} and {\sc Xenon}100~\cite{Xenon10010} 
are testing particle physics models in which DM is coupled to the Higgs boson or in which the DM interactions with matter occur only at loop level.

Although the CDMS, EDELWEISS and {\sc Xenon}100 searches for the DM have not given a discovery  so far, 
 there is a long-standing claim by DAMA/LIBRA~\cite{DAMA} that
sees an $8\sigma$ evidence for an annual modulation in the nuclear recoil signal of low-mass DM. Recently this claim received some support from the 
results of CoGeNT experiment~\cite{CoGeNT}. In order to reconcile this result with the negative searches from other experiments, un-orthodox DM scenarios such as 
Inelastic Dark Matter (iDM)~\cite{IDM, IDM2} have been put forward. Testing all those scenarios is of utmost importance for understanding the nature of DM.

Recently the {\sc Xenon}100 experiment published new data after 101 days of data taking that
 probes values of the spin-independent DM/nucleus cross section down to
$\sigma_{\rm SI}\approx 7.0\times 10^{-45}\,{\rm cm}^2$~\cite{Xenon100}.
They observed 3 signal candidate events with the expected background of  $(1.8\pm 0.6)$ events. This result leads to the most 
stringent limit on DM interactions today, and further constrains the best motivated DM models.

In this paper we perform a fit to the new {\sc Xenon}100 data and apply the results to constrain several model dependent as well as model independent  
scenarios of DM. First, motivated by the DAMA/LIBRA anomaly we study  whether the iDM, as an explanation to the observed annual modulation,
can survive the new {\sc Xenon}100 results. We find that this is not the case and the DAMA/LIBRA anomaly needs another explanation. After that we study
the simplest possible DM model, the scalar singlet model. 

However, most of our effort goes to the studies of SUSY models. We study the Constrained
Minimal Supersymmetric Standard Model (CMSSM) as well as the low energy phenomenological MSSM (pMSSM). In doing that we apply also constraints 
on SUSY parameter space coming from the LHC at $\sqrt{s} = 7\TeV$  with 1.1/fb of integrated luminosity,  
new data from D0, CMS and LHCb on $B_s\to \mu^+\mu^-$, and new data from Tevatron on the top quark mass~\cite{CMS,ATLAS,Bsmumu}. 
We find that the {\sc Xenon}100 data stringently constrains the ``well-tempered" neutralino
scenario~\cite{welltempered}, and excludes the corresponding ``focus-point'' parameter region of the CMSSM~\cite{Feng:1999mn}.
Therefore we perform a model independent analysis of that scenario. We derive constraints of new {\sc Xenon}100 data on the CMSSM and pMSSM
parameter spaces and study the implications of those to the expected results in the LHC experiments. We find that, after EPS 2011 conference,
the ``Higgs-resonance" mechanism of DM freeze-out is completely excluded. The allowed CMSSM parameter space is pushed towards higher sparticle masses, the allowed range of $\tan\beta$ is constrained from below and from above, and the fine tuning of the parameters has become more severe. 
We show that the DM direct detection experiments and the LHC are complementary in constraining the supersymmetric parameter space.

The paper is organized as follows. In section ~2 we perform the fit of new {\sc Xenon}100 data and apply this to iDM in section~3.
In section~\ref{scalar} we study the scalar singlet model.
In section~\ref{CMSSM} we study supersymmetric dark matter. We conclude in section~6.

\section{Fit of {\sc Xenon}100 data}

We perform a fit of the new data relased by the {\sc Xenon}100 collaboration \cite{Xenon100}, taken between January and June 2010, which corresponds (before any cut) to an exposure of 101\,days$\,\times\,$48\,kg. {\sc Xenon}100 reports 3 events in the signal region which are consistent with an estimated background of $1.8\pm 0.6$ events.

For fixed values of $M_{\rm DM}$ and $\sigma_{\rm SI}$, the number of signal events expected in a given direct-detection experiment depends on astrophysical, nuclear and experimental parameters which are more or less well-measured.
We build a global $\chi^2$ including these extra parameters as nuisances and marginalize over them at the end. %Practically one has to make some reasonable choice to reduce the amount of computer-time needed.

\begin{table}[t]
\begin{center}
\begin{tabular}{c|cc}
parameter & range & distribution\\ \hline
$v_0~ {\rm km\,s^{-1}}$&200 - 240 & Gaussian: $220\pm 10$\\
$v_{\rm esc}~ {\rm km\,s^{-1}}$&498 - 618 & see \cite{RAVEsurvey}\\
$\mathcal L_{\rm eff}$&$\pm 2\sigma$& see \cite{Xenon100}
\end{tabular}
\end{center}
\caption{\label{paramfit}\em The parameters ranges for the {\sc Xenon}100 fits.}
\end{table}

On the astrophysical side one has to know both the local DM density, $\rho_\odot $, and the DM velocity distribution. In the standard halo model the latter is fixed to be a Maxwellian distribution with a given r.m.s.\ velocity $v_0$, truncated at certain escape velocity $v_{E}$:
\begin{equation}
\frac{dN}{d\vec v}\equiv f(\vec v)\propto e^{-v^2/v_0^2}\theta(v-v_E).
\end{equation}
The preferred values of the two velocities are $v_0=220~{\rm km\,s^{-1}}$ and $v_E=544~{\rm km\,s^{-1}}$~\cite{RAVEsurvey}.
We assume that $v_0$ has a Gaussian uncertainty of $\pm 10~{\rm km\,s^{-1}}$, and we allow for a $2\sigma$ variation of it in the fit. 
We take from~ \cite{RAVEsurvey} the probability distribution of the $v_E$ parameter. 
The local DM density $\rho_\odot$ is kept fixed to $0.3~{\rm GeV\,/cm^{3}}$ since the effect of its variation is 
just equivalent to an overall rescaling of $\sigma_{\rm SI}$ (the quantity really probed by direct detection experiments is their product
$\sigma_{\rm SI}\rho_\odot$).

In terms of this set of parameters the differential scattering rate per unit detector mass is
\begin{equation}
\label{rate}
\frac{d R}{d E_{\rm nr}}=N_T\frac{\rho_\odot}{M_{\rm DM}}\int_{|\vec v|>v_{\min}} \!\!\!d^3 v~ v \,f(\vec v)\, \frac{d\sigma_{{\rm DM}~ N}}{d E_{\rm nr}},
\end{equation}
where $N_T$ is the number of nuclei in the target per unit detector mass, $M_{\rm DM}$ is the DM mass and $v_{\min}$ is the minimal velocity which allows a $n$uclear $r$ecoil with energy $E_{\rm nr}$. $\sigma_{{\rm DM}~ N}$ is the nucleus-DM cross section.

We will be interested in spin-independent processes. Nuclear physics enters into the determination of form factors which modulates the cross section according to the momentum transfer:
\begin{equation}
\frac{d \sigma_{{\rm DM}~ N}}{d E_{\rm nr}}=\frac{m_N\sigma_{\rm SI}}{2 v^2\mu_n^2}\frac{\left(f_p Z+f_n(A-Z)\right)^2}{f_n^2} F^2(E_{\rm nr}).
\end{equation}
Here $m_N$ is the nucleus mass, $\mu_n$ is the nucleon-DM reduced mass, and $\sigma_{\rm SI}$ is the spin-independent nucleon-DM cross section. We fix our choice of $F$ to the standard Helm form factor \cite{helmff}. We furthermore assume equal coupling strength to both protons and neutrons: $f_p\approx f_n=1$.

The last important source of uncertainty comes from the experiment itself. In {\sc Xenon}100 the energy of a recoiling nucleus is inferred from the number of recorded photo-electrons (called \emph{S1 signal}) emitted after the prompt relaxation of the excited nucleus. The relation between S1 and the recoil energy is given by
\begin{equation}
S1(E_{\rm nr})=3.6~{\rm PE}\times E_{\rm nr}\times \mathcal L_{\rm eff}(E_{\rm nr}).
\end{equation}
The value of the function $\mathcal L_{\rm eff}$ at recoil energies (nuclear) smaller than 5\,keV is not measured and must be extrapolated. Such extrapolation is crucial to put limits on light dark matter: the larger $\mathcal L_{\rm eff}$ is, the stronger the limits are. We use the recent $\mathcal L_{\rm eff}$ measurement
 from \cite{Xenon100}, based on \cite{Plante:2011hw}, and we allow in our fit $2\sigma$ excursions from its central value.

In view of the small number of observed events 
we perform an event-by-event fit, such that we construct the $\chi$-square~\cite{chisupernova}
\begin{equation}
\hat\chi^2_{\rm Xenon100}(M_{\rm DM}, \sigma_{\rm SI},\{p_i\})=\hat\chi^2_{\{p_i\}}(M_{\rm DM}, \sigma_{\rm SI})+\sum_i\chi^2_i(p_i),
\end{equation}
where $\chi^2_i$ is the one associated to the parameters themselves and
$\hat\chi^2_{\{p_i\}}$ is the $\chi$-square for fixed values of the parameters $p_i$ 
($v_0$, $v_{\rm esc}$ and $\mathcal L_{\rm eff}$),
evaluated in terms of the S1 signal along the lines of \cite{Xenon10010}. We thus include both a Poissonian smearing of the signal and a Gaussian smearing of the energy measurement. The events are collected in the window $4\,{\rm PE}\leq {\rm S1}\leq 30\,{\rm PE}$. We use the acceptance from \cite{Xenon100} and we model the background as a flat distribution in the signal region normalized to reproduce the expected number of background events.

We then marginalize $\hat\chi^2_{\rm Xenon100}$ with respect to the nuisance parameters $p_i$ defining
\begin{equation}
\chi^2_{\rm Xenon100}(M_{\rm DM}, \sigma_{\rm SI})=\min_{\{p_i\}} \hat\chi^2_{\rm Xenon100}(M_{\rm DM}, \sigma_{\rm SI},\{p_i\}),
\end{equation}
where the minimum is taken in the range summarized in Table~\ref{paramfit}.

\section{Inelastic Dark Matter}
\begin{figure}[t]
$$\includegraphics[width=0.5\textwidth]{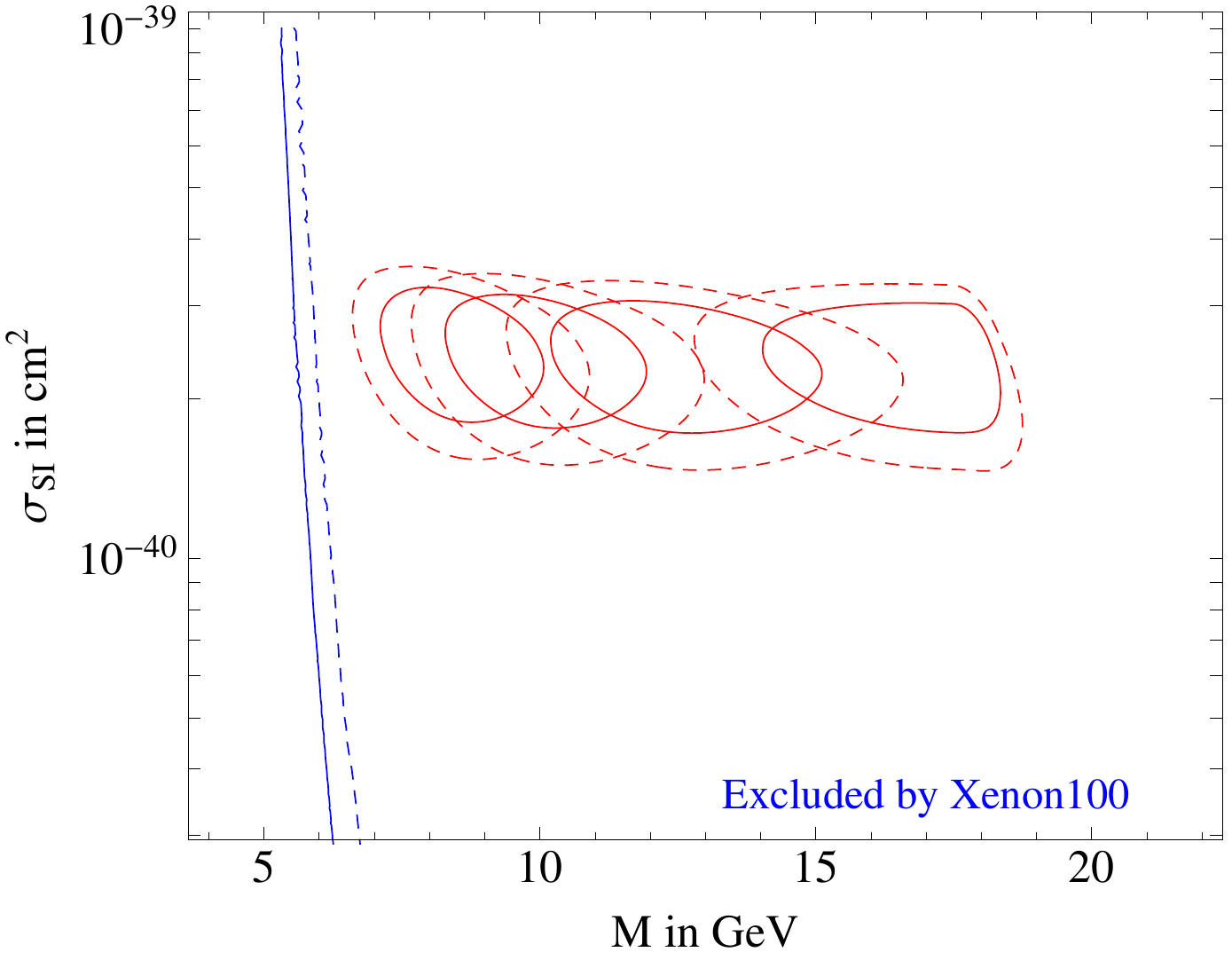}$$
\caption{\em The 95, 99.7\% confidence level contours  for 2 d.o.f. for the DAMA modulated signal under the assumption of elastic scattering on sodium atoms. We assume different values of the sodium quenching factor, 0.2, 0.3, 0.4 and 0.5, from right to left. The blue lines are the {\sc Xenon}100 exclusion curves at 95\% (continuous curve), 99.7\% (dashed) confidence level. We assume $v_0=220\, {\rm km\,s^{-1}}$ and $v_{esc}=544\,{\rm km\,s^{-1}}$ for the DAMA fit. We neglect channeling.
\label{damalm}}
\end{figure}

The DAMA collaboration reported an 8$\sigma$ evidence for an annual modulation of their nuclear recoil signal \cite{DAMA} which is compatible with light dark matter scattering on nuclei. Such signal conflicts with other direct detection experiments and 
(as already remarked by the {\sc Xenon}100 Collaboration~\cite{Xenon100})
is strongly disfavored by {\sc Xenon}100 under the assumption of an elastic interaction of the dark matter with the Na nuclei of the DAMA detector. We show this in Fig.~\ref{damalm} where we plot the region favored by elastic scattering on sodium at DAMA together with the {\sc Xenon}100 exclusions.
The DAMA confidence levels are obtained by a simple $\chi^2$ fit of the first 12 bin of fig.~9 in~\cite{DAMA}, 
showing the energy dependence of the seasonal modulation.
%showing the measured modulated rate.

The strong bounds already from previous experiments, prompted theorists to consider
Inelastic Dark Matter~\cite{IDM} (iDM), namely the idea that the DM detection process could be
${\rm DM}~N \to {\rm DM}' \; N$ where the new state ${\rm DM}'$ is heavier than the DM state by an amount $\delta=M_{{\rm DM}'}-M_{\rm DM}$.
It was pointed out that the modified kinematics
could make the DAMA anomaly compatible with experiments, such as CDMS, that use lighter nuclei.
Indeed the minimum DM velocity needed to scatter on a nucleus with mass $m_N$ giving recoil energy $E_{\rm nr}$ is~\cite{IDM}
\beq v_{\rm min} > \sqrt{\frac{1}{2m_N E_{\rm nr}}} \left(\frac{m_N E_{\rm nr}}{\mu_N} + \delta\right),\eeq
where $\mu_N$ is the reduced mass of the DM/nucleus. 
For the mechanism to be effective the inelasticity mass splitting is assumed to be 
around $  100\,{\rm keV}$, which is the order of the typical nuclear recoil energy.
By assuming progressively smaller tails of the uncertain DM velocity it is possible to avoid progressively stronger constraints
from experiments with light nuclei. For the same reason it is possible to enhance the relative modulation in the signal claimed by DAMA.

\begin{figure}[t]
$$\includegraphics[width=1\textwidth]{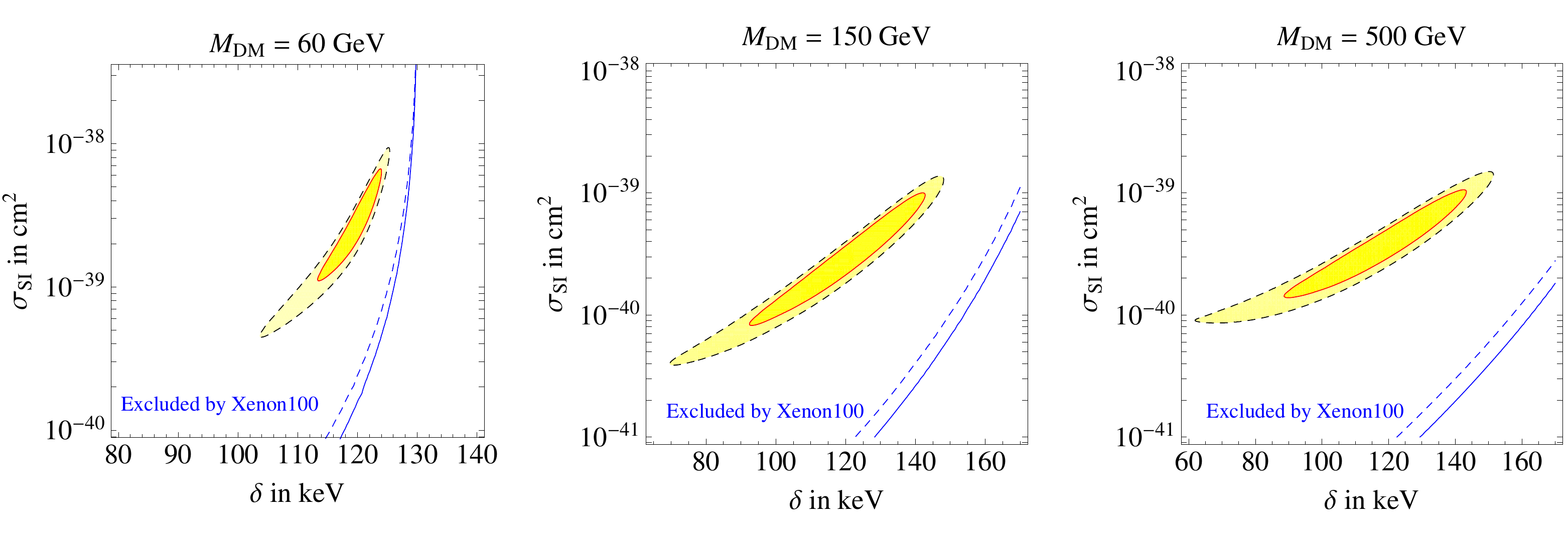}$$
\caption{\em The 95, 99.7\% confidence level contours  for 2 d.o.f. for iDM fit to DAMA, together with the  95, 99.7\% exclusion curves from {\sc Xenon}100 data (full and dashed respectively). We fix the iodine quenching factor to 0.085. Sodium quenching and channeling are irrelevant for these values of the mass.
\label{iDM}}
\end{figure}

Xenon nuclei have a mass very close to iodine which is, at DAMA, the dominant source of recoils for DM particles with mass above $\mathcal O(10\, {\rm GeV})$. This fact, together with the high exposure of {\sc Xenon}100 allows to put strong constraints on the iDM hypothesis. This is shown in Fig.~\ref{iDM} where we plot 95\% and 99.7\% confidence level contours for the parameters favored by DAMA in the $(\delta, \sigma_{\rm SI})$ plane, together with the {\sc Xenon}100 exclusion curves. 
The $v_0$, $v_{\rm esc}$ and $\mathcal L_{\rm eff}$ parameters have been here fixed to their median values. 
We find that the iDM hypothesis to explain DAMA is disfavoured throughout its parameter space. This conclusion is sound, independent of the various uncertainties thanks to the similarity between Xe and I nuclei. This can be understood in a simple way \cite{IDM2}. The rate observed by DAMA can be written as
\begin{equation}
\mathcal B+\mathcal S(1+ a \cos \omega(t-t_0)),
\end{equation}
where $\mathcal B$ is a time-independent unknown background, $\mathcal S$ is the mean DM signal and $a$ describe the size of the modulation. $t_0=$2 June, is where the peak of the modulation observed by DAMA is located. The size $a\mathcal S$ of the measured modulation is $\sim 0.02 \,{\rm counts\,kg^{-1}\,day^{-1}}$ in the range $2\,{\rm keV}\leq E_{\rm nr}\leq 6\,{\rm keV}$. Due to the iodine quenching factor, $q_I\approx 0.085$, this region translates to $23\,{\rm keV}\leq E_{\rm nr}\leq 70\,{\rm keV}$ at Xenon. In the region of overlap between the former range and the acceptance window $8.4-44.6$ keV, the DM scattering rate at Xenon is given by
\begin{equation}\label{ratexe}
R_{\rm Xenon100}\sim A\times \epsilon\times \mathcal S\left(1/a+\cos \omega(t-t_0)\right).
\end{equation}
$\epsilon$ and $A$ are respectively the efficiency of the {\sc Xenon}100 cuts and its acceptance. Together they amounts to roughly $0.2$. Integrating eq.~(\ref{ratexe}) over the energy range and over the time of Xenon data-taking one finds
\begin{equation}
N_{\rm Xenon100}> \mathcal O(50) {\rm~ events},
\end{equation}
where we folded in the exposure and we used $a<1$. It is thus clear where the exclusion curves shown in Fig.~\ref{iDM} comes from.
 %We do not discuss the possibility to rescue the iDM hypothesis or the DAMA result with solid-state effects such as channeling.

\begin{figure}
$$\includegraphics[width=0.5\textwidth]{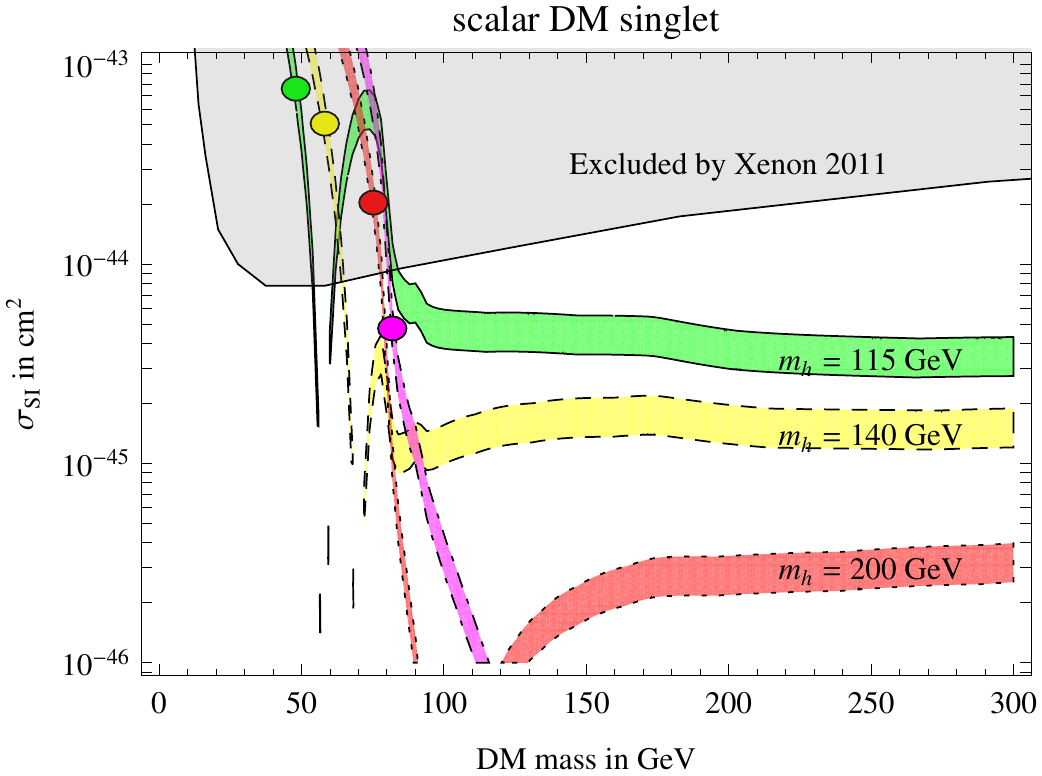}$$
\caption{\em Predictions of the scalar singlet model for a few values of the Higgs boson mass: 115 GeV (green), 140 GeV (yellow),
200 GeV (red), 300 GeV (magenta).
The dots are the predictions of the constrained model of~\cite{Kadastik:2009dj,CDMSth}.
\label{fig:singlet}}
\end{figure}

\section{Scalar singlet Dark Matter model}\label{scalar}
We consider a DM model obtained adding to the Standard Model a Dark Matter real singlet scalar field $S$
coupled to the Higgs doublet $H$ as described by the following Lagrangian~\cite{McDonald, Burgess, Barger,DMS}:
\beq \Lag =\Lag_{\rm SM } + \frac{(\partial_\mu S)^2}{2} - \frac{m^2}{2}S^2
-\lambda S^2 |H|^2\,, \label{eq:L}\eeq
 invariant under the matter parity $S\to -S,$ the discrete gauge symmetry of scalars carrying $B-L$~\cite{Kadastik:2009dj}.
$S$ is the DM field and its mass is given by $M_{\rm DM}^2 = m^2 +\lambda V^2$ with $V =246\GeV$.
So the model has 2 free parameters, $M_{\rm DM}$ and $\lambda$.
Assuming that the relic DM abundance equals its cosmologically measured value the model is able of predicting
the relation in the plane ($M_{\rm DM},\sigma_{\rm SI}$) plotted in Fig.\fig{singlet}.
Given that the Higgs boson mass is also presently unknown, we plotted such relation for a few values of the Higgs mass: 115 GeV (green, favored by precision data), 140 GeV (yellow, compatible with precision data),
200 GeV (red, disfavored by precision data); 300 GeV (magenta, strongly disfavored by precision data).

The very small $\sigma_{\rm SI}$ predicted around $M_{\rm DM}=m_h/2$ is due to the Higgs resonance enhancement of the
cosmological DM annihilation rate.
This quantity is computed as in~\cite{CDMSth}, already including the 3 body final states whose relevance was emphasized in~\cite{3}.

In absence of a theoretical motivation for having $m$ comparable to the Higgs mass, \cite{Kadastik:2009dj,CDMSth} considered the case $m=0$, such that the model has one parameter less
and is able of predicting a point  in the plane ($M_{\rm DM},\sigma_{\rm SI}$).
Such prediction is also shown in Fig.\fig{singlet}, for the same values of the Higgs boson mass.

 We remark two uncertainties not explicit from the plot.
First, the {\sc Xenon}100 exclusion bound is plotted assuming $\rho_\odot = 0.3\GeV/{\rm cm}^3$ for the local DM density.
This is the canonical value routinely adopted in the literature, with a typical associated error bar of $\pm 0.1$ GeV/cm$^{3}$. Recent computations  found a higher central value closer to $0.4\GeV/{\rm cm}^3$~\cite{Nesti}
that would imply stronger bounds on the cross section $\sigma_{\rm SI}$.

Second,  the prediction for the conventional spin-independent DM/nucleon cross section is:
\beq\label{eq:f} \sigma_{\rm SI}%=\frac{|{\cal A}|^2}{16\pi s}
= \frac{\lambda^2 m_N^4 f^2}{\pi M_{\rm DM}^2 m_h^4},\ \eeq
where $f$ parameterizes the nucleon matrix element:
\beq \langle{N} | m_q \bar{q}q | N\rangle\equiv  f_q m_N [\bar N N] ,\qquad
f = \sum_{q=\{u,d,s,c,b,t\}}\!\!\!\!\! f_q = \frac{2}{9}+\frac{5}{9}\sum_{q=\{u,d,s\}}\!\! f_q.  \eeq
The main uncertainty comes from $f_s$.
The recent analyses use $f= 0.56\pm 0.11$~\cite{Ellis},
or $f= 0.30\pm0.015$~\cite{lattice}, in agreement with the lattice results~\cite{Toussaint:2009pz} and phenomenological determination~\cite{Koch:1982pu}.
Here and in the following we assume the default value in the {\sc Micromegas} code: $f=0.467$~\cite{micromegas}.

\section{Supersymmetry}\label{CMSSM}
In this section we study the impact of new {\sc Xenon}100 data on constraining SUSY models. 
We first consider the CMSSM, the most popular SUSY  model with an unified scalar mass $m_0,$ 
an unified gaugino mass $M_{1/2}$ and an unified  trilinear scalar $A$-term at the GUT scale. 
Given that {\sc Xenon}100 adds to many other experimental constraints, we perform a global fit to all relevant data 
as described in the next subsection. Most importantly, we include the recent CMS and ATLAS constraints on
the CMSSM parameters space that are based on the LHC data with 1.1/fb presented in july 2011.
Our fits extend the previous ones obtained in the similar studies in~\cite{Feldman:2011me,Allanach:2011ut,Buchmueller:2011aa,Bechtle:2011dm,Akula:2011zq,Conley:2011nn,Akula:2011dd} without EPS 2011 data.
%As a result we are going to show the complementarity between 
%the {\sc Xenon}100 and LHC results in constraining CMSSM and study the implications of the {\sc Xenon}100 results for the LHC 2011 run.

One of the results of our research is that the  ``well-tempered" neutralino scenario is stringently constrained
by the {\sc Xenon}100 new result. Motivated by that we perform a generic analyses of the ``well-tempered" neutralino 
and show that, quite model independently, this scenario is now stringently constrained.

Finally we relax the CMSSM constraints on the particle spectrum coming from the unification relations and 
study a generic low energy phenomenological MSSM, the pMSSM. We identify the generic {\sc Xenon}100 constraints on pMSSM.

\subsection{Global fit of supersymmetric models}

Global fits of the SUSY models have recently been performed  by several groups in the context of LHC studies, and our results agree with them.  
Therefore we focus on the impact of the new {\sc Xenon}100 data in constraining SUSY models.
Here, we briefly describe our procedure, and the differences with respect to previous approaches.

\begin{table}[t]
\begin{center}
\begin{tabular}{c|cc}
quantity & experiment & Standard Model\\ \hline
$\alpha_3(M_Z)$~\cite{alpha3} &  $0.1184 \pm 0.0007$ & parameter\\
$m_t$~\cite{topmass} & $173.1\pm 0.9$&parameter\\ 
$m_b$~\cite{PDG} & $4.19\pm 0.12$&parameter\\ \hline
$\Omega_{\rm DM} h^2$~\cite{OmegaDM} & $0.112 \pm 0.0056$ & 0\\
$\delta a_\mu$~\cite{Davier:2010nc} & $(2.8\pm0.8)10^{-9}$ & 0\\
BR$(B_d\to X_s\gamma)$~\cite{bsg} & $(3.50\pm0.17)\,10^{-4}$ & $(3.15\pm0.23)\,10^{-4}$ \\
BR$(B_s\to \mu^+\mu^-)$~\cite{Bsmumu} & $(0.9\pm0.6)\,10^{-8}$& $(0.33\pm 0.03)\,10^{-8}$\\
BR$(B_u\to \tau\bar\nu)$/SM~\cite{Buchmueller:2009fn} & $1.25\pm0.40$ & 1\\
\end{tabular}
\end{center}
\caption{\label{tab:1}\em The data we fit, together with LHC and {\sc Xenon}100 bounds.}
\end{table}

We perform a random scan of the parameter space and calculate the sparticle spectrum as well as DM relic abundance 
using the {\sc Micromegas} public code~\cite{micromegas}.
We vary the relevant input SM parameters ($m_t$, $\alpha_3$, $m_b$) within $5\sigma$ experimental errors assuming a Gaussian distribution. 
The CMSSM parameters ($m_0$, $M_{1/2}$, $A_0$, $\tan\beta$, ${\rm sign}\,(\mu)$) are generated randomly in the ranges $(m_0, M_{1/2})\sim (0,4000)$~GeV,
$A_0\sim (-3 m_0, 3 m_0),$ $\tan\beta\sim(1,60)$ and ${\rm sign}(\mu)=\pm1$ with a flat distribution.
Using  {\sc Micromegas} we compute all 
observables for each sampling point and select those rare cases that reproduce all experimental data within $5\sigma$ errors.

We do not use any special technique such as Markov chains, which makes the scanning more efficient,
but that might lead to missing some local best-fit regions.
Using the computing power of the Baltic Grid we get about $200,000$ good points that satisfy all the experimental criteria.
For such points we compute the global $\chi^2$ using the data summarized in Table~\ref{tab:1}:
\beq 
\chi^2 = \chi^2_{\rm SM~parameters}+\chi^2_{\rm observables} + \chi^2_{\rm LHC} + \chi^2_{\rm {Xenon}100} \, .
\eeq
Notice that we do not include electroweak precision observables, 
that do not have a significant impact on the result, and that
are not well approximated simply by the oblique parameters ~\cite{Shap}.

In this work we perform a purely phenomenological fit, considering values of the sparticles masses
significantly above the weak scale (up to $4\TeV$),
ignoring the  theoretical  issue of naturalness (see~\cite{AS,Cassel:2011tg} for a recent analysis).
Technically, this is achieved as follows: when plotting the $\chi^2$ as function of one or two parameters, we minimize it with respect to all other parameters.
The fit is mainly driven by the DM abundance and by the apparent anomaly in the anomalous magnetic moment of the muon, and agrees with the 
fits in~\cite{Buchmueller:2011aa,Bechtle:2011dm,Buchmueller:2010ai}.
Given that it might not be a real anomaly, we also show regions at relatively high
confidence levels.

We keep the nuclear matrix elements and the DM local density fixed to their default values in  {\sc Micromegas}, as already discussed in the previous section.

\begin{figure}[t]
$$\includegraphics[width=0.45\textwidth]{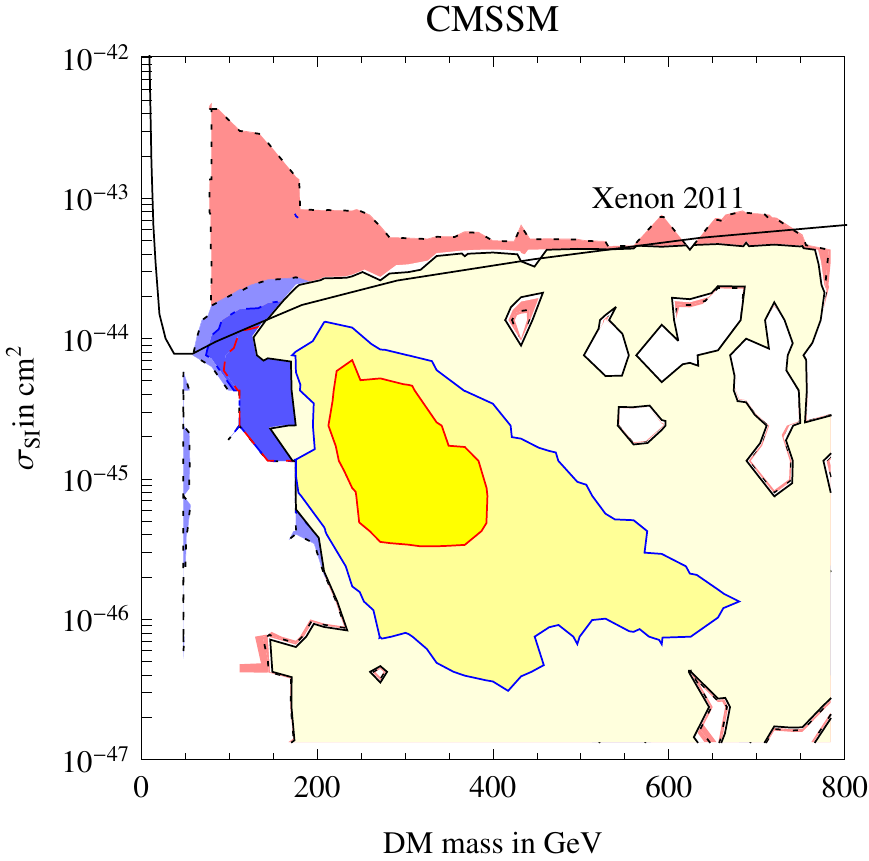}\qquad\includegraphics[width=0.45\textwidth]{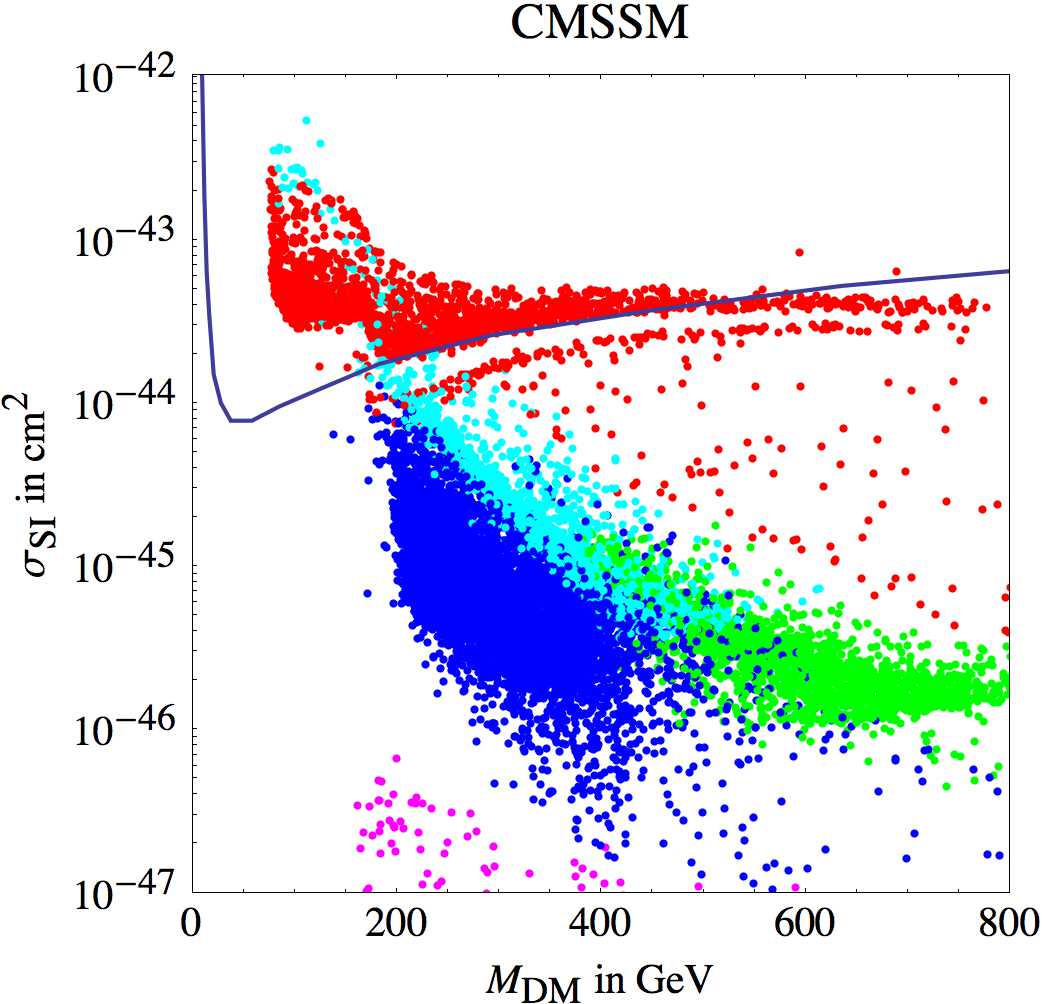}$$
\caption{\em {The ($M_{\rm DM},\sigma_{\rm SI}$) plane in the CMSSM}.
In the left panel we show the global fit:
the yellow regions surrounded by continuous contours are the best fit including the {\sc Xenon}100 and LHC data, at
$68,95,99.7\%$ confidence levels for 2 d.o.f.  The red (blue) regions surrounded by dashed contours are the corresponding regions now excluded by {\sc Xenon}100 (LHC).
In the right panel we show points with $\Delta\chi^2<4^2$, colored according to the DM annihilation mechanism.
The red dots in the upper region excluded by the {\sc Xenon}100 correspond to the ``well-tempered" neutralino, green via the heavy Higgs resonance, cyan via neutral Higgses with $\tan\beta$-enhanced couplings,
blue via slepton co-annihilations, magenta via stop co-annihilations.
\label{fig:CMSSMfits3}}
\end{figure}

\subsection{The CMSSM results}

Fig.\fig{CMSSMfits3}a shows our global CMSSM fit for the DM mass $M_{\rm DM}$ and spin-independent DM-nucleus cross section $\sigma_{\rm SI}$
measured by {\sc Xenon}100 experiment.
The yellow regions surrounded by continuous contours are the best fit regions including the {\sc Xenon}100 and LHC data, at
$1,2$ and $3\sigma$ level ($68,95,99.7\%$ confidence levels for 2 d.o.f).
We also show, as red regions surrounded by dashed contours, the previous best-fit regions at the same
confidence levels now excluded by {\sc Xenon}100 at more than $3\sigma$.
Obviously, such excluded regions lie around the {\sc Xenon}100 exclusion bound at $90\%$ confidence level (the continuous curve in the figure).

\begin{figure}[t]
$$\includegraphics[width=0.45\textwidth]{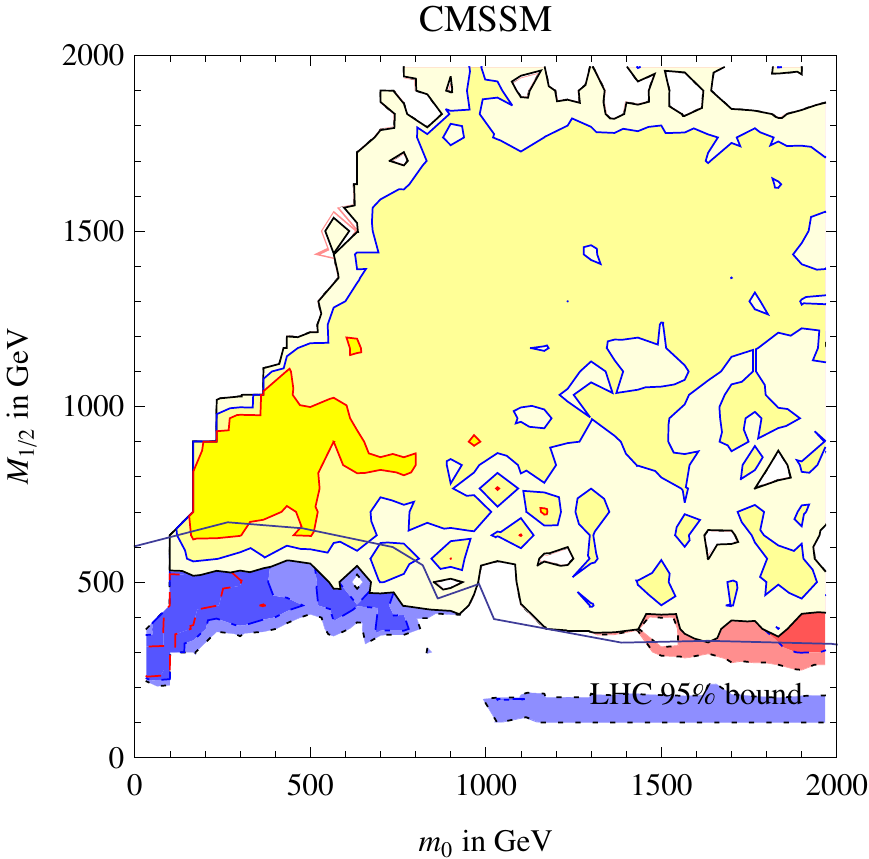} \qquad \includegraphics[width=0.45\textwidth]{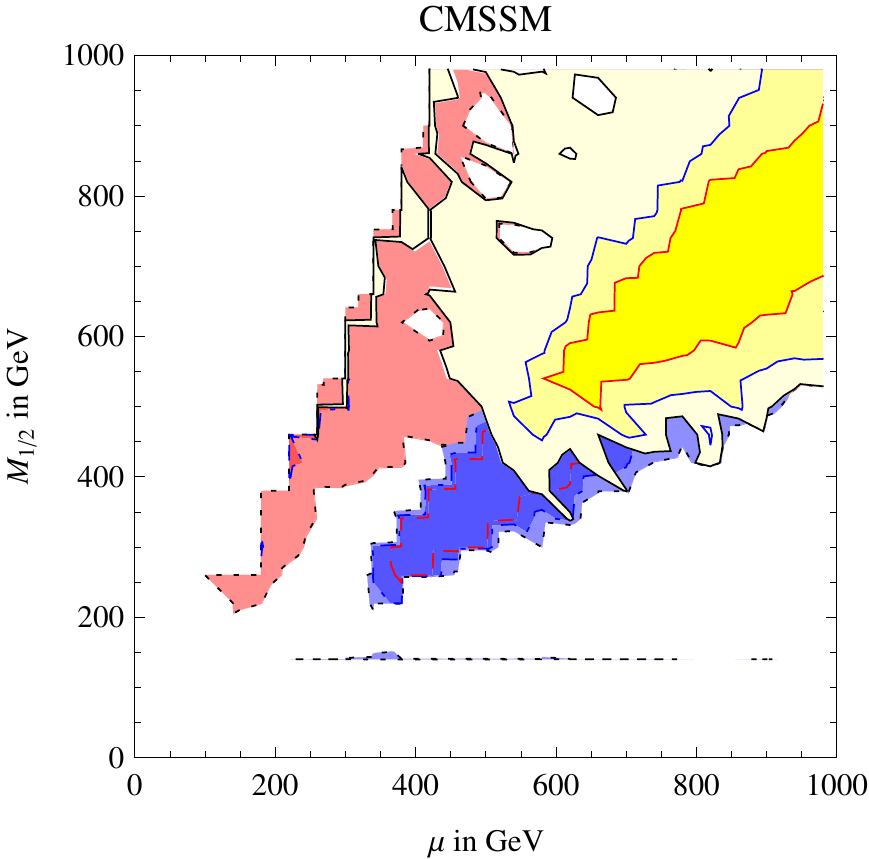}$$
\caption{\em {Global CMSSM fits in $(m_0, M_{1/2})$ (left) and in $(\mu, M_{1/2})$ (right) planes}. 
The red (blue) contours are excluded by {\sc Xenon}100 (LHC), other details are as in Fig.\fig{CMSSMfits3}a.
\label{fig:CMSSMfits2}}
\end{figure}

Within the CMSSM, thermal freeze-out of
neutralino DM can reproduce
the observed DM  cosmological abundance
according to a few qualitatively distinct mechanisms,
that correspond to different fine-tunings.
To interpret this result we therefore discriminate such distinct cases, plotting in
Fig.\fig{CMSSMfits3}b the points of the CMSSM parameter space
(also imposing a reasonably good global fit, $\Delta\chi^2<4^2$)
colored according to their dominant DM annihilation mechanisms.
We identify the following distinctive regions:
\begin{enumerate}
\item Red points (``arm'' in the upper part of the plot)
have $|\mu|\approx M_1$ such that
the lightest neutralino has a significant Higgsino component
(``well-tempered" neutralino~\cite{welltempered}).
In the CMSSM this happens in the fine-tuned region with large $m_0$ and small $\mu$ in the so-called ``focus-point" region~\cite{Feng:1999mn}.
The double structure of the  ``arm" in Fig.\fig{CMSSMfits3}b shows that both signs of the $\mu$
parameter can give the correct DM relic abundance while $\mu>0$ is favored by the $(g-2)_\mu$ anomaly.
Comparison with Fig.\fig{CMSSMfits3}a shows that the new {\sc Xenon}100 data excludes or disfavors
this scenario.
\item Blue points (around the center of the plot) have $m_{\tilde{\ell}}\approx M_{\rm DM}$,
corresponding to  slepton co-annihilations. They provide the best global fit.

\item Green points (around the bottom-right part of the figure)
have  $M_{\rm DM}\approx m_A/2$,
corresponding to the resonant DM annihilations mediated by the heavy CP-odd Higgs boson $A$.
The light-Higgs resonance (that arises for  $M_{\rm DM}\approx m_h/2$ and consequently implies
a relatively light gluino) has now been excluded by the new  bounds on the gluino mass obtained
by the CMS and ATLAS collaborations after $1.1/$fb of data.

\item Cyan points represent parameters in which one moves away from  Higgs resonances
(we arbitrarily assume a $10\%$ off-degeneracy to discriminate  from the previous case) but the
DM annihilation processes are still mediated via  all neutral Higgses in $s$-channel.
A large enough Higgs couplings to the SM fermions and to DM is obtained thanks to
$\tan\beta>45$ and to a non-negligible Higgsino component of the DM neutralino.
Consequently some of these points have a large direct detection cross section.

\item Magenta points (at very low $\sigma_{\rm SI}$) correspond to stop co-annihilations. The latter region is/will be further constrained by the LHC
data that presently excludes light stops.
\end{enumerate}
We remark that, within the CMSSM, the LEP and LHC bounds have excluded one more mechanism that was considered 
more plausible because its non-fine-tuned nature: the bino annihilations via light slepton exchange in $t$-channel.
In the light of our results the experimental data favours slepton co-annihilation as the mechanism for generating neutralino thermal relic abundance.
This result is a consequence of the fact that the fit is largely dominated by the measurement of muon anomalous magnetic moment that requires
relatively light sleptons.

\begin{figure}[t]
$$\includegraphics[width=0.45\textwidth]{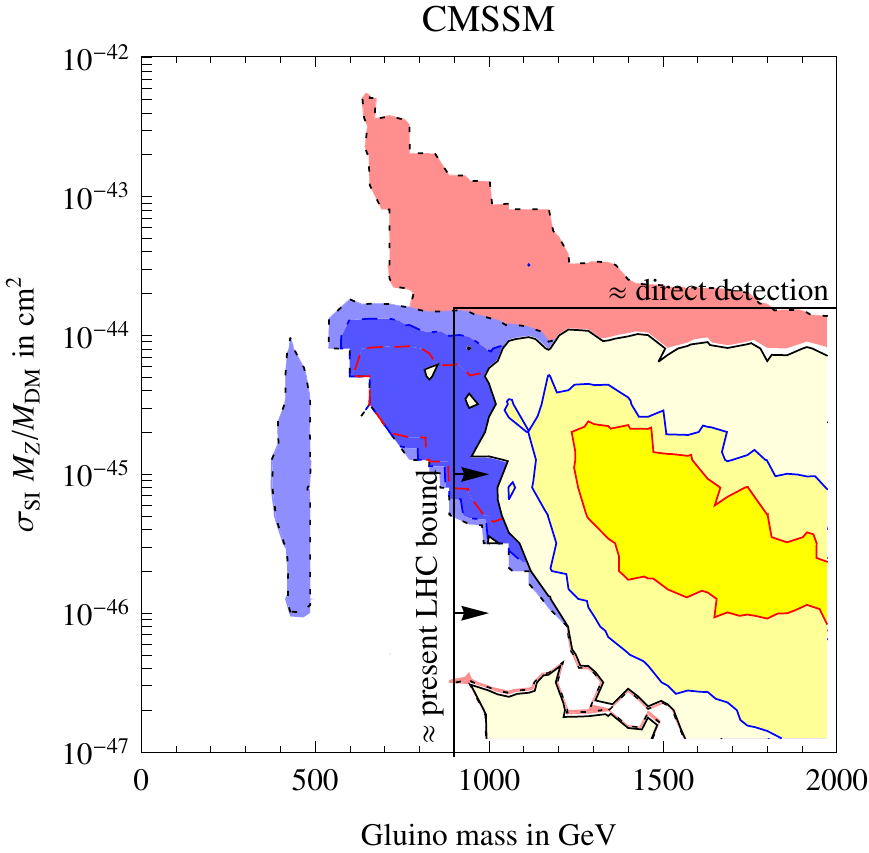}\qquad \includegraphics[width=0.45\textwidth]{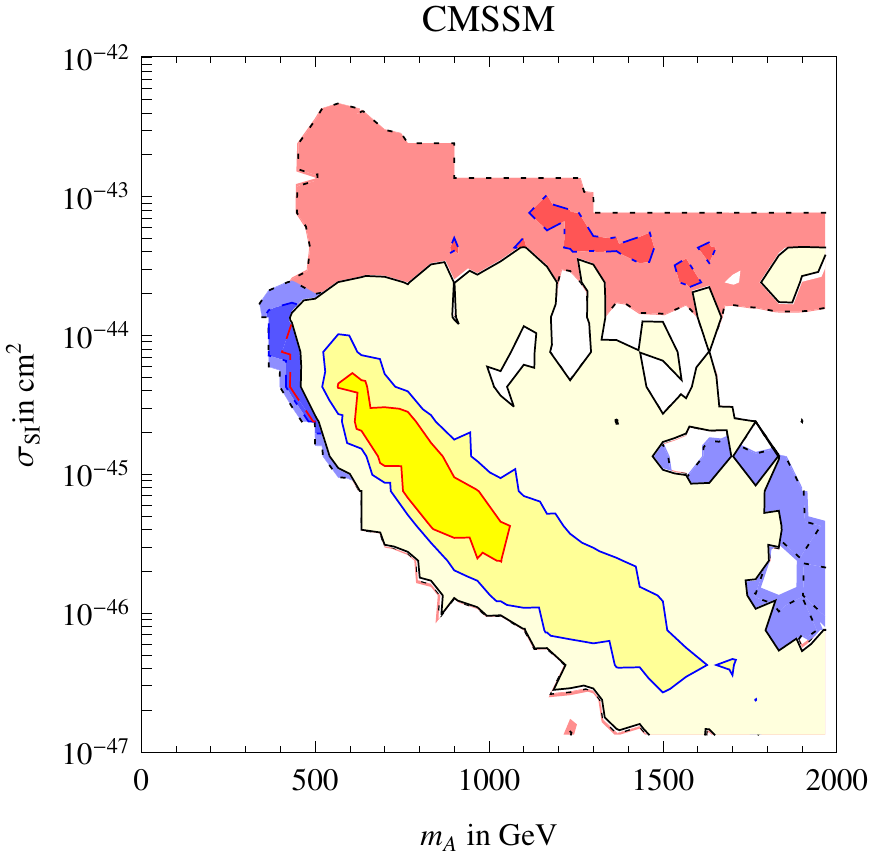}$$
\caption{\em {Global CMSSM fits in ($M_{\tilde g},\sigma_{\rm SI}$)  (left) and in ($m_{\rm A},\sigma_{\rm SI}$) (right) planes}. 
The red (blue) contours are excluded by {\sc Xenon}100 (LHC), other details are as in Fig.\fig{CMSSMfits3}a.
\label{fig:CMSSMfits2p}}
\end{figure}

In the light of these considerations, we can now discuss the global CMSSM fits for other parameters.
Fig.\fig{CMSSMfits2} shows some 2-dimensional fits as indicated in the figure.
We see that {\sc Xenon}100 strongly disfavors regions with small $|\mu|$ and partly also with small $M_{1/2}$ 
(the very small allowed region at $3\sigma$ level for small $M_{1/2}$ below the red excluded region corresponds to the Higgs resonance DM annihilation mechanism).
Within the CMSSM a small $|\mu|$ can be realized in the ``focus-point" scenario that requires a multi-TeV $m_0$.
Such a region was previously not favored by global fits and is now further disfavored by the {\sc Xenon}100 bound that disfavors
a neutralino DM with a significant higgsino component. The fact that small values of $|\mu|$ below 400~GeV are now excluded
by {\sc Xenon}100 data is perhaps one of the most important results of our work.

\begin{figure}[t]
$$\includegraphics[width=\textwidth]{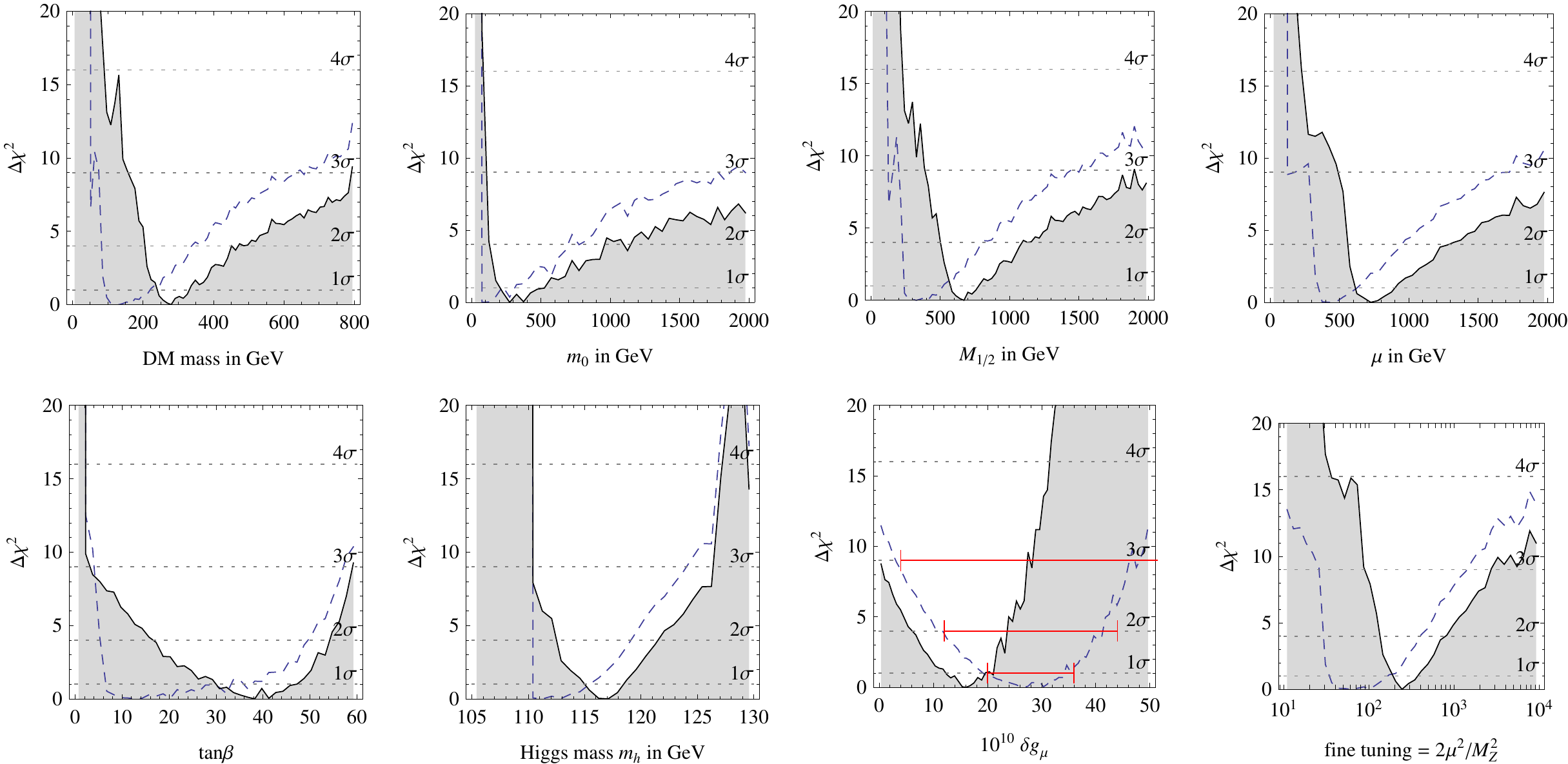}$$
\caption{\em Global CMSSM fit before (dashed curve) and after (continuous curve) the {\sc Xenon}100 and LHC results.
In the upper  panels we fit the DM mass (left) the $m_0$ (middle left), $M_{1/2}$ (middle right), $\mu$ (right) parameters;
in the lower  panels we fit $\tan\beta$ (left), the Higgs mass (middle left), the muon $g-2$ (middle right) and the fine-tuning parameter (right).
\label{fig:CMSSMfits1}}
\end{figure}

Fig.\fig{CMSSMfits2p}a compares the LHC prospects (here characterized by the approximate gluino mass reach),
with the prospects of direct detection experiments
(characterized by $\sigma_{\rm SI} M_{\rm DM}/M_Z$, which is the only parameter probed by experiments when $M_{\rm DM}$
is much heavier than the target nucleus).
We see that LHC excluded the
Higgs resonance region, that within the CMSSM necessarily has a light gluino with mass $M_3\approx 400\GeV$,
and reached the best-fit region. On the other hand, Fig.\fig{CMSSMfits2p}b shows that, after {\sc Xenon}100, there is a clear correlation between
the spin-independent DM direct detection cross section and $m_{\rm A}.$  Although the {\sc Xenon}100 results exclude only a very small region
of the parameter space at low values of   $m_{\rm A},$ this implies that the CMSSM charged Higgs boson mass must exceed $M_{H^+}> 400$~GeV
and is inaccessible at the 7~TeV LHC.

Finally, Fig.\fig{CMSSMfits1} shows our fits for the DM mass and for a few key parameters: $m_0$, $M_{1/2}$,
$\tan\beta$, $m_h$ and $2\mu^2/M_Z^2$ (which is a simple measure of fine-tuning, closely related to the $\mu$ term).
In all cases new data had a significant impact: the exclusion of lighter sparticles, moves the best fit to higher sparticle masses 
and consequently to higher $\tan\beta$ in order to fit the $g-2$ anomaly.
As shown in the $g-2$ panel, there starts to be some tension between its experimental measurement (error bands)
and the CMSSM predictions.
Adding the LHC and Xenon100 bounds, the global $\chi^2$ at the best-fit point worsens by $\Delta \chi^2 \approx 4.3$.
Furthermore, the lower right panel shows that the fine-tuning grows up to the few hundred range (for a dedicated discussion see the updated version of~\cite{AS}).

%In the light of that result LHC upgrade to the nominal 14~TeV center of mass energy is needed to cover the CMSSM best fit region. 

%
\begin{figure}
$$\includegraphics[width=0.5\textwidth]{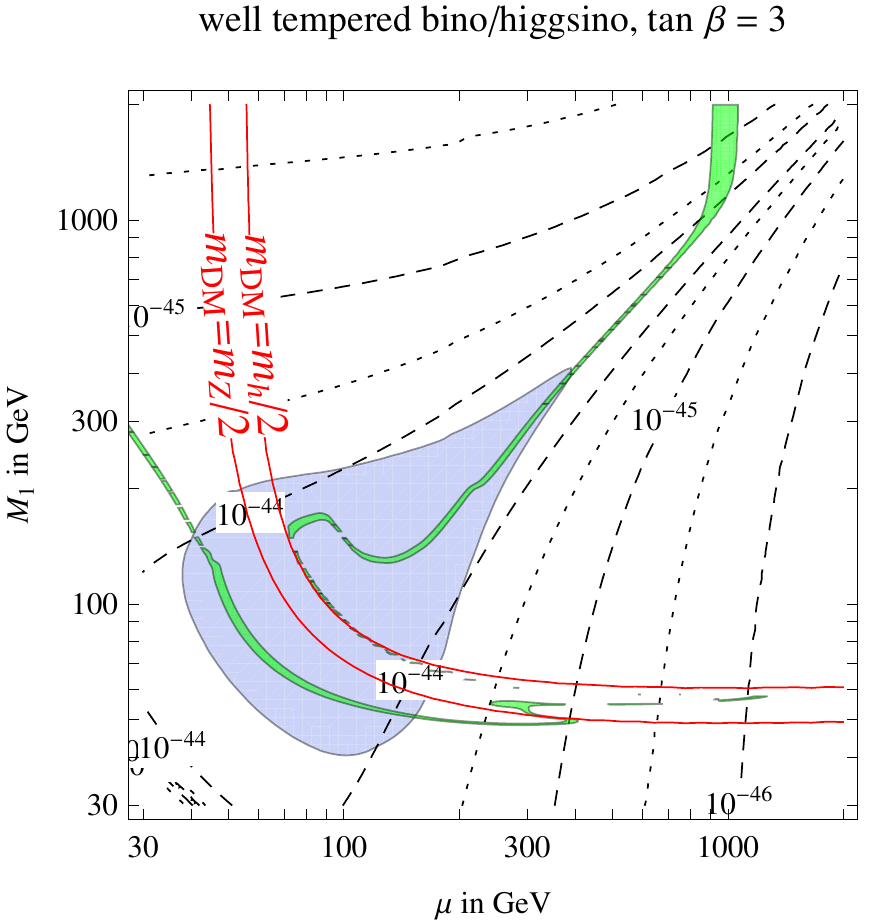}\qquad
\includegraphics[width=0.5\textwidth]{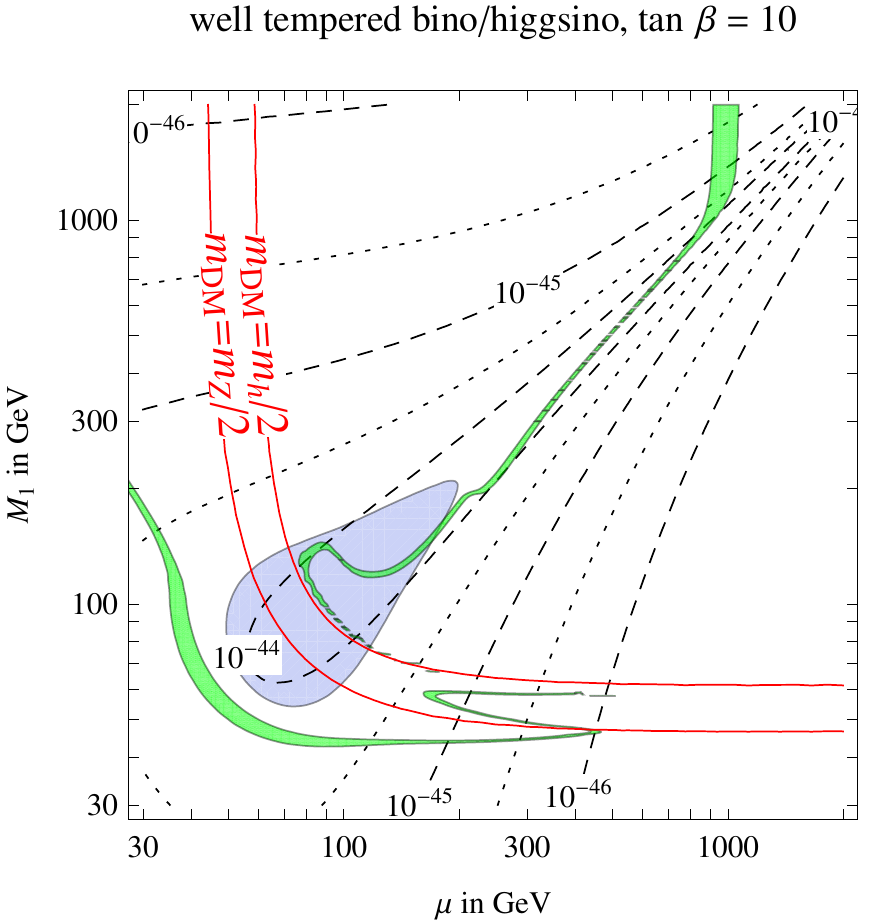}$$
\caption{\em Model independent ``well-tempered" neutralino scenario. The $3\sigma$ range for the cosmological DM abundance is reproduced within the
green strip.  The gray region is excluded by {\sc Xenon}100~\cite{Xenon100}.
\label{fig:WT}}
\end{figure}

\subsection{``Well-tempered" neutralino}\label{welltempered}
As we have shown, within the CMSSM, the {\sc Xenon}100 result has interesting implications on the ``well-tempered"  neutralino~\cite{welltempered}.
Here we study this scenario of DM generation in SUSY models in a model-independent way.
Within the MSSM, the neutralino is a mixed state of bino, wino, higgsino.
None of them, in a pure state, allows thermal DM with a weak scale mass $M_{\rm DM}\approx M_Z$:
\begin{itemize}
\item The pure higgsino couples to the $Z$ too strongly, such that for $M_{\rm DM}\approx M_Z$ its thermal abundance is too low (the cosmological abundance is obtained for a heavy $M_{\rm DM}\approx 1\TeV$ higgsino); furthermore, for the same reason it is experimentally excluded by direct searches.

\item
The pure wino similarly has too much co-annihilations with charged winos, such that for $M_{\rm DM}\approx M_Z$ its thermal abundance is too low (the cosmological abundance is obtained for a heavier $M_{\rm DM}\approx 2.7\TeV$ taking into account electroweak Sommerfeld effects~\cite{Hisano}).
Contrary to the previous case,  having no coupling to the $Z$ it is allowed by direct searches.

\item
The pure bino, instead, has no couplings and no co-annihilations, such that its cosmological abundance would be too high.
\end{itemize}
Given that the bino has opposite problems with respect to the higgsino or the wino,
it is possible to find a good DM candidate by appropriately mixing them~\cite{welltempered}.
A mixed bino/wino still has no couplings to the $Z$, such that it is not interesting for direct detection; furthermore
it requires $M_1\approx M_2$ at the weak scale and is not compatible with unification of gaugino masses,
$M_1\approx M_2\approx M_3$ at the GUT scale.

We thereby focus on a mixed bino/higgsino.
In the limit where we can ignore all other heavier sparticles, its phenomenology is fully described by 3 parameters: the bino mass term $M_1$, 
the higgsino mass term $\mu$ (we assume them to be positive)
and $\tan\beta$.
The observed thermal relic DM abundance is reproduced in the green strip in Fig.\fig{WT}
(left panel for $\tan\beta=3$ and right panel for $\tan\beta=10$).
The region with $M_1\approx\mu\approx M_Z$ was allowed, but its large direct detection cross section is now disfavored by {\sc Xenon}100
(gray region).   An improvement of the {\sc Xenon}100 bound by a factor of few would fully exclude the whole ``well-tempered" neutralino scenario,
unless the local DM density or the nuclear matrix element $f$ of eq.\eq{f} are significantly lower than what is assumed in our  computation.

The minor tilt at $M_{\rm DM}\approx m_t$ is due to the top quark threshold.
At lower masses, the cosmologically allowed region of Fig.\fig{WT} is affected by the $Z$ and Higgs resonances ($2M_{\rm DM} = M_Z$ or $m_h$ respectively, indicated as red curves).
At larger masses, the ``well-tempered" neutralino region terminates at $\mu\approx 1\TeV$, where the (almost) pure higgsino becomes a good DM candidate.

%$$\begin{array}{c|ccc}
%& \hbox{pure bino} & \hbox{pure higgsino} & \hbox{pure wino}\\ \hline
%\hbox{$\Omega = \Omega_{\rm DM}$  for $m=$} & 0 & 1.0\TeV & 2.7\TeV\\
%\hbox{coupling to the $Z$} & 0 & 0 & g_2
%\end{array}
%$$

\begin{figure}[t]
$$\includegraphics[width=0.45\textwidth]{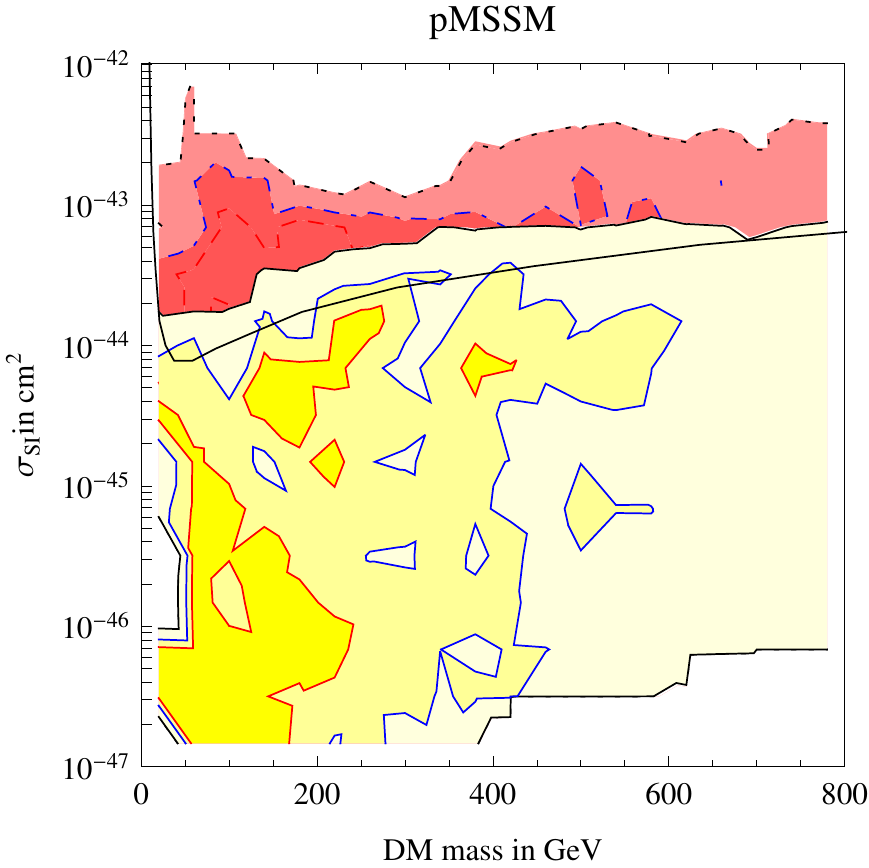}\qquad\includegraphics[width=0.45\textwidth]{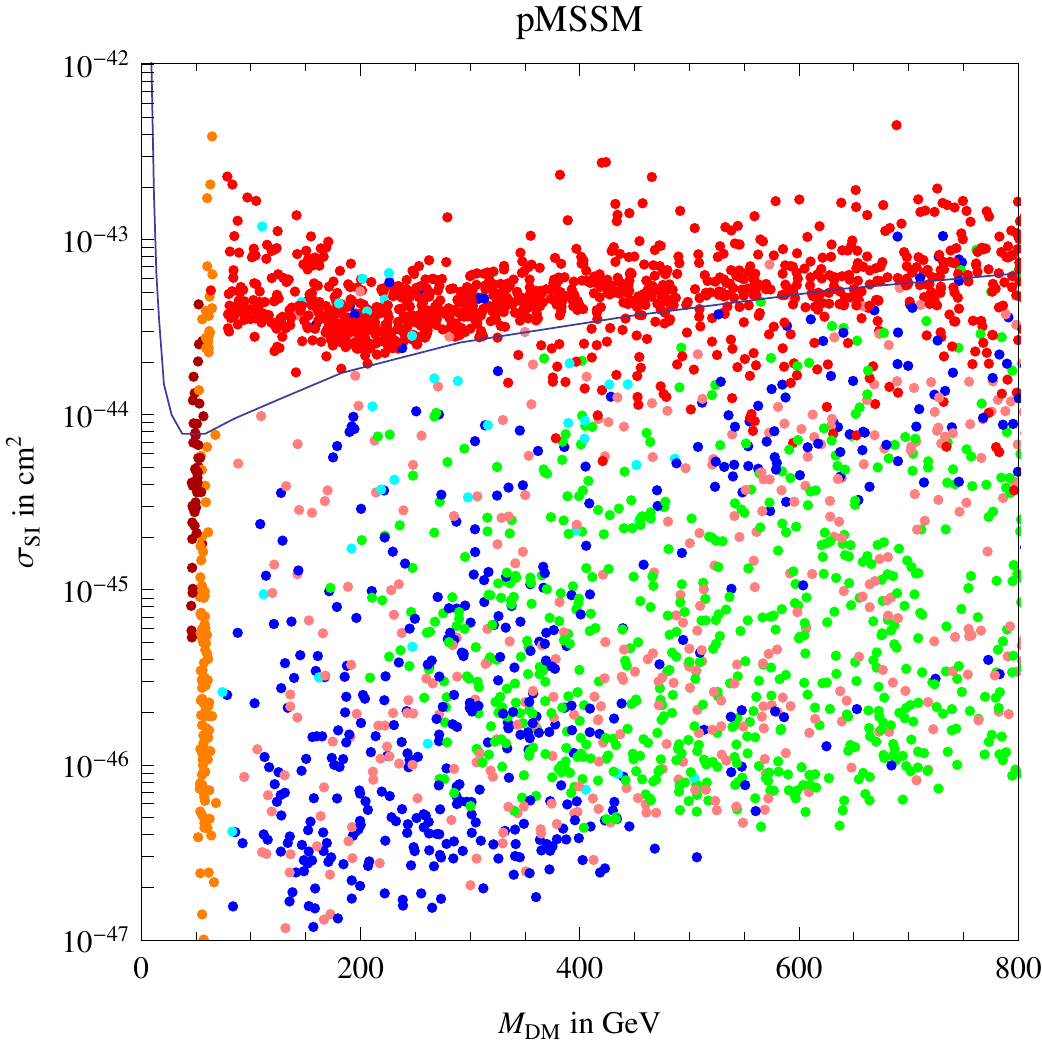}$$
\caption{\em  The pMSSM fit (left panel) in $(M_{\rm DM},\sigma_{\rm SI})$ plane and the corresponding DM generation mechanisms (right panel) in
the colour code analogous to Fig.\fig{CMSSMfits3} except for three new regions: the pink dots denote ``well-tempered" bino/wino; 
the dark red dots the $Z$-resonance at $M_{\rm DM}\approx M_Z/2$; the orange dots denote the light-higgs resonance at $M_{\rm DM}\approx m_h/2$.
\label{fig:pMSSM}}
\end{figure}

\subsection{Supersymmetry beyond the CMSSM: pMSSM}\label{pMSSM}
In order to relax the theoretical constraints of the CMSSM on the SUSY mass spectrum, we also performed a similar analysis 
for free low energy SUSY parameters of the MSSM. The low energy phenomenological MSSM, pMSSM, is characterized by
the three gaugino masses $M_1, M_2, M_3,$ the higgsino mass $\mu,$ the common squark masses $m_{\tilde{q}}$,
the left and right-handed slepton masses $m_{\tilde{L}}, m_{\tilde{E}}$, the Higgs mass parameter $m_A$ and by  $\tan\beta$. We randomly scan all these parameters.

Fig.\fig{pMSSM} shows our results, in the same notations as the corresponding CMSSM figure, Fig.\fig{CMSSMfits3}.
Because the squark and slepton masses are not related any more via the GUT relations, light sleptons become available for a good fit of $(g-2)_\mu$ that dominates
the fit. At the same time, heavy squarks can suppress any new contributions to $b\to s \gamma.$
Therefore a wider range of parameters becomes allowed by the global fit, with still a preference for relatively light DM mass,
that again mostly corresponds to neutralino/slepton coannihilations (blue dots in Fig.\fig{pMSSM}b).
Again the region disfavored by the new {\sc Xenon}100 data mostly corresponds to the ``well-tempered" bino/higgsino  (upper red dots in Fig.\fig{pMSSM}b).
Two new regions appear: ``well-tempered" bino/wino (pink dots, at any mass with $\sigma_{\rm SI}$ below the {\sc Xenon}100 bound),
and $Z$-resonance (dark red dots at $M_{\rm DM}\circa{>}M_Z/2$).
There are now significant overlaps among the various kinds of DM neutralinos and the mechanisms for generating the DM relic abundance 
are not well separated. Although the pMSSM parameter space is more complicated than the CMSSM one, the results are qualitatively similar.

\section{Conclusions}

We have performed a fit to the new {\sc Xenon}100 data and, based on that, analyzed several well motivated DM scenarios.
We show that {\sc Xenon}100 disfavors the Inelastic Dark Matter interpretation and the light Dark Matter interpretation
of the DAMA/LIBRA claim as well as other hints for light DM by CoGeNT and CDMS II. 

\medskip

The first results of the {\sc Xenon}100 experiment exclude part of the parameter space of Dark Matter models coupled to the
Higgs boson. The constrained version of the scalar singlet dark matter model (where the DM mass is predicted in terms
of DM coupling to the Higgs boson), previously favored by the CDMS hint, gets disfavored.
Such model now needs an independent mass term for the scalar singlet around the weak scale,
with its associated naturalness problem.

\medskip

In the context of supersymmetry, we show model independently that the {\sc Xenon}100 data 
disfavors the generic scenario of  ``well-tempered'' bino/higgsino  proposed as a way to get neutralino DM around the weak scale.
Within the CMSSM, the {\sc Xenon}100 data excludes small $|\mu|$, small $m_{\rm A}$ regions of the parameter space
previously allowed in the global fit at $3\sigma$ level. As a result the ``focus-point" region  is ruled out and,
most likely, the CMSSM charged Higgs boson mass is not accessible at 7~TeV LHC. 
Furthermore the ``Higgs resonance'' mechanism for neutralino annihilation, which
implies $M_{\rm DM}\approx m_h/2$ and consequently a relatively light gluino, has now been probed and excluded
by LHC data taken during 2011.

The remaining CMSSM best fit parameter space at $3\sigma$ level is relatively compact in which the DM relic abundance is 
generated by the slepton co-annihilation processes.  
In view of the new LHC bounds, the CMSSM best fit region moves to larger DM mass and consequently to
lower direct detection cross section. The remaining CMSSM parameter space is more fine tuned than before.

%Today neither the {\sc Xenon}100 nor the  LHC experiments probe the best fit CMSSM parameter space yet
%but the sensitivity of  both experiments will enter into this region.

We  obtained qualitatively similar constraints also for the pMSSM, where there is no connection between masses of
colored and uncolored sparticles.  As a consequence the light Higgs resonance is still allowed.

\small

\begin{figure}[t]
$$\hspace{-0.5cm}\includegraphics[width=0.32\textwidth]{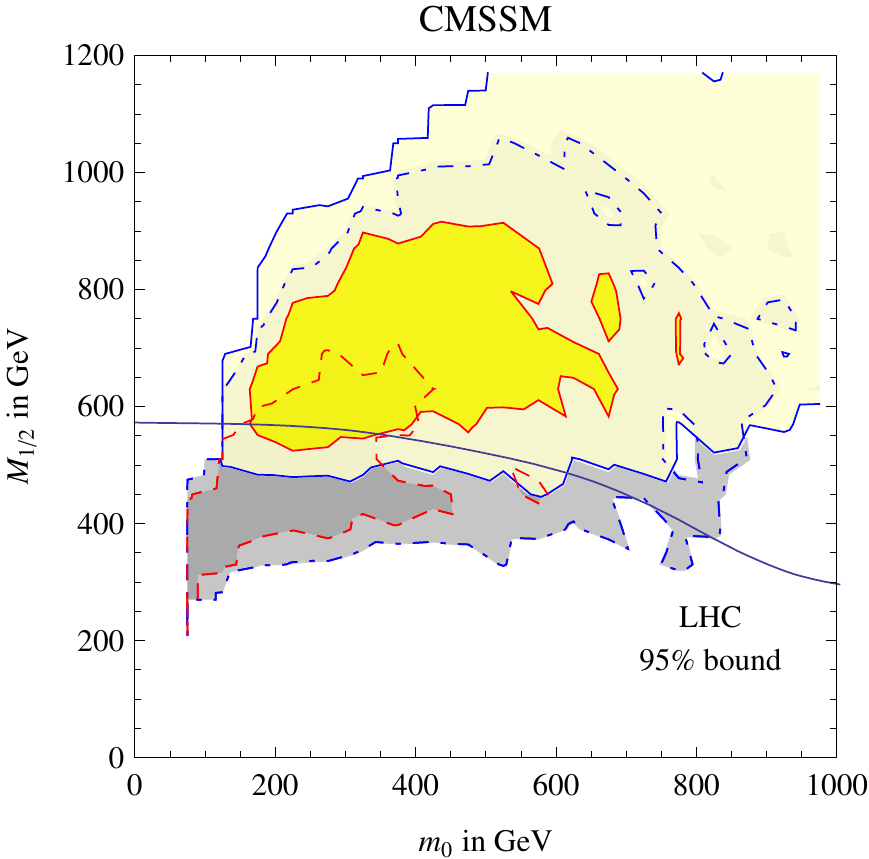}\quad
\includegraphics[width=0.32\textwidth]{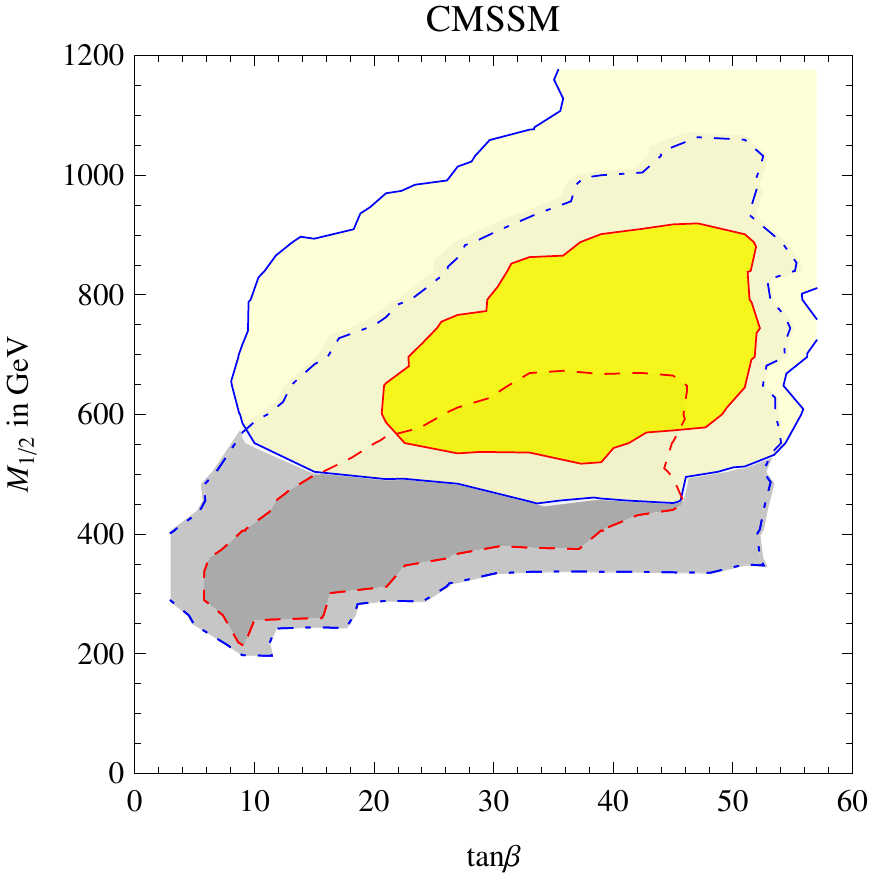}\quad
\includegraphics[width=0.32\textwidth]{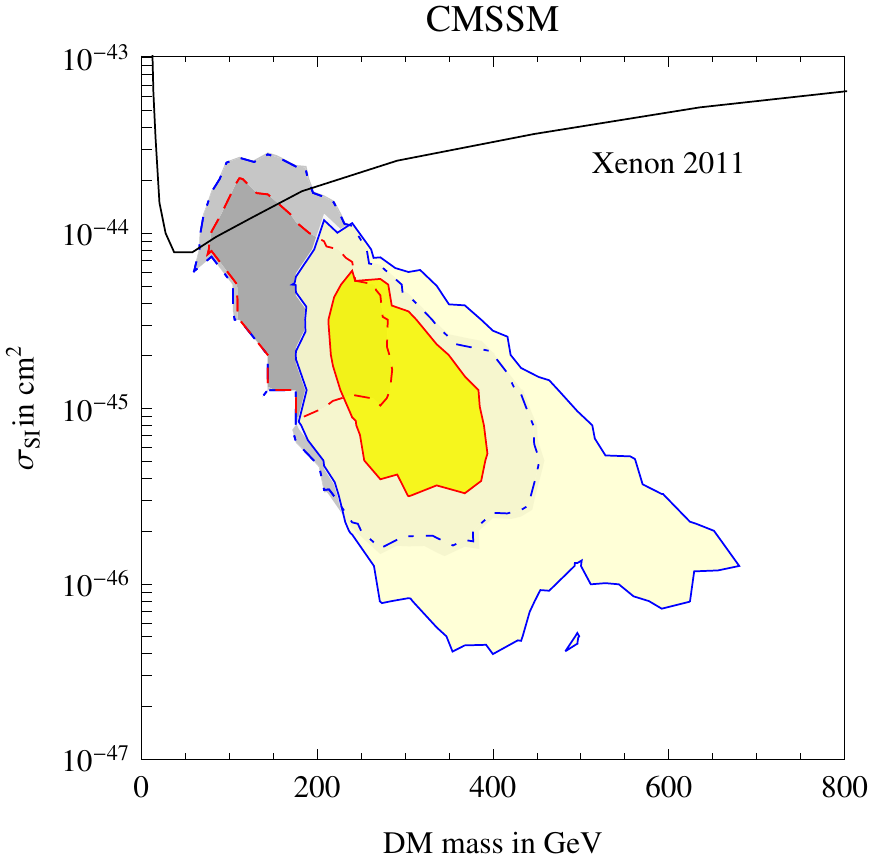}$$
\caption{\em Impact of the latest {\sc Xenon}100, Tevatron and LHC data with {\rm 1.1/fb} on CMSSM global fits.
In yellow the present global fit at $68$ and $95\%$ C.L. for 2 dof; in gray the previous result without such data.
Fig.\fig{CMSSMnews}a shows that LHC bounds move the best fit region to higher sparticle masses;
then a higher $\tan\beta$ is  needed to fit the $(g-2)_\mu$ (see Fig.\fig{CMSSMnews}b), 
and a lower direct detection cross section is obtained (see Fig.\fig{CMSSMnews}c). 
The slepton co-annihilation mechanism of DM freeze-out is favored;
while now both the ``well-tempered'' bino/higgsino mechanism as well as the ``Higgs resonance'' mechanism
are no longer allowed by the global fit.
\label{fig:CMSSMnews}}
\end{figure}

\paragraph{Note added (27/7/2011).}  
The paper has been fully updated at the light of the new results presented at the EPS-HEP-2011 conference (21--27 july 2011),
including data from ATLAS and CMS on supersymmetric particles and about MSSM Higgs bosons, 
new data from D0, CMS and LHCb about $B_s\to \mu^+\mu^-$, and new data from Tevatron on the top quark mass.
All those results are available in the web site \url{http://eps-hep2011.eu}.
The new figure Fig.\fig{CMSSMnews} shows the impact of present {\sc Xenon}100 and LHC data on CMSSM global fits.
The allowed  CMSSM parameter space has moved to considerably large values of the mass parameters $m_0$ and $M_{1/2}$. 
As a result, the Higgs resonance region of DM freeze-out is now completely ruled out.
The allowed values of $\tan\beta$ are constrained both from below (due to larger sparticle masses, larger $\tan\beta$ is needed to
fit the muon $g-2$ anomaly) and from above (due to the new constraint on BR$(B_s\to \mu\mu)$, mainly from LHCb). 
The DM spin independent direct detection cross section is now predicted to be a factor of few below the present {\sc Xenon}100 bound.
As seen in Fig.\fig{CMSSMfits1}, the overall fit to the CMSSM parameters has become worse and fine tuning of the CMSSM parameters  has further increased.

%When this articles was under review, ATLAS experiment published new preliminary results for searches of supersymmetry in the 
%context of CMSSM~\cite{ATLAS2}.
%To exemplify the effect of new ATLAS data, we present in  Fig.\fig{CMSSM_ATLAS}  exactly the same as in Fig.\fig{CMSSMfits3}a but performing fits with 
%the new LHC data. 
%Comparison of Fig.\fig{CMSSM_ATLAS}  and Fig.\fig{CMSSMfits3}a shows that new data restricts CMSSM more than before but qualitatively
%the two plots are in agreement and our results do not change qualitatively.

\paragraph{Note added (18/4/2011).} After our fits to the new {\sc Xenon}100 data were performed and studies of iDM completed, a similar dedicated 
study appeared by the {\sc Xenon} Collaboration itself \cite{Xenon100iDM}. 
Our results agree in that the iDM is strongly disfavoured as an explanation to the  DAMA/LIBRA annual modulation signal.

\paragraph{Acknowledgements}
We thank Alexander Pukhov for valuable communication and for providing us with a new version of Micromegas package.
This work was supported by the ESF grants  8090, 8499, MTT8 and by SF0690030s09 project. The work of D.P. is supported by the Swiss National Science Foundation under contract No. 200021-116372. The work of M.F. is supported in part by the European Programme ``Unification in the LHC Era",  contract PITN-GA-2009-237920 (UNILHC).

%

%\end{multicols}

\newpage

\normalsize

\section*{Addendum: 2012 data}\label{10in}
\begin{figure}
$$\includegraphics[width=0.5\textwidth]{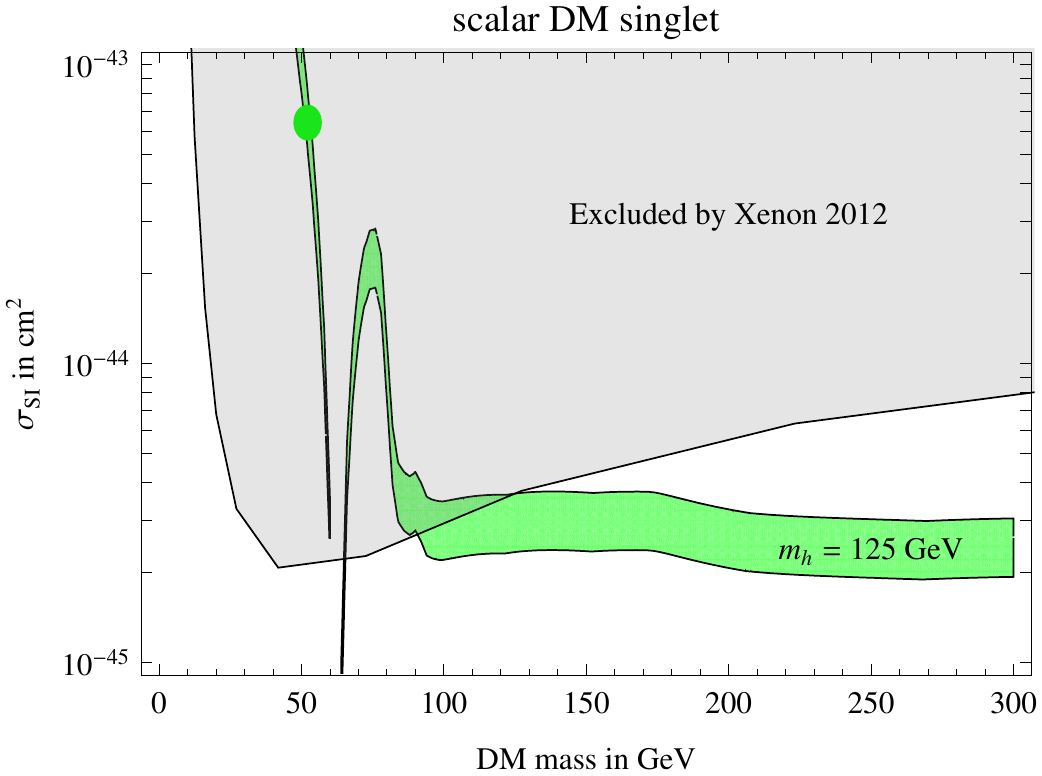}$$
\caption{\em Update of fig.\fig{singlet}.
Predictions of the scalar singlet model.
\label{fig:singlet2012}}
\end{figure}

We update our previous results including
the new data released during July 2012 by the {\sc Xenon100} collaboration~\cite{Xenon2012}
(sensitivity to $\sigma_{\rm SI}$ improved by a factor of 5, 
see fig.\fig{singlet2012}) and by the ATLAS and CMS collaborations (stronger bounds on sparticle masses, and
the measurement of the Higgs mass: $m_h =125.5\pm0.5\GeV$~\cite{higgs}).

Fig.\fig{singlet2012} shows 
that the scalar singlet DM model considered in section~\ref{scalar}
now survives only if  DM is heavier than about 100 GeV.
In particular this excludes the sub-case where all the DM mass comes from the Higgs vacuum expectation value
(green dot in fig.\fig{singlet2012}).

\medskip

\begin{figure}[p]
$$\includegraphics[width=0.5\textwidth]{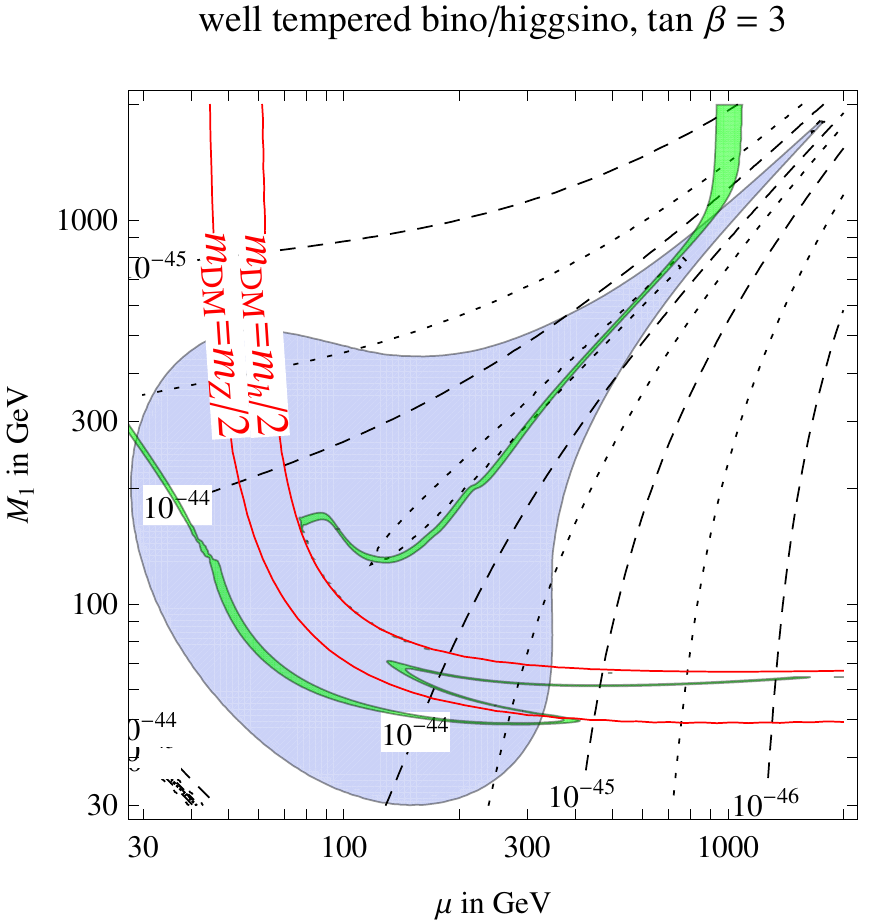}\qquad
\includegraphics[width=0.5\textwidth]{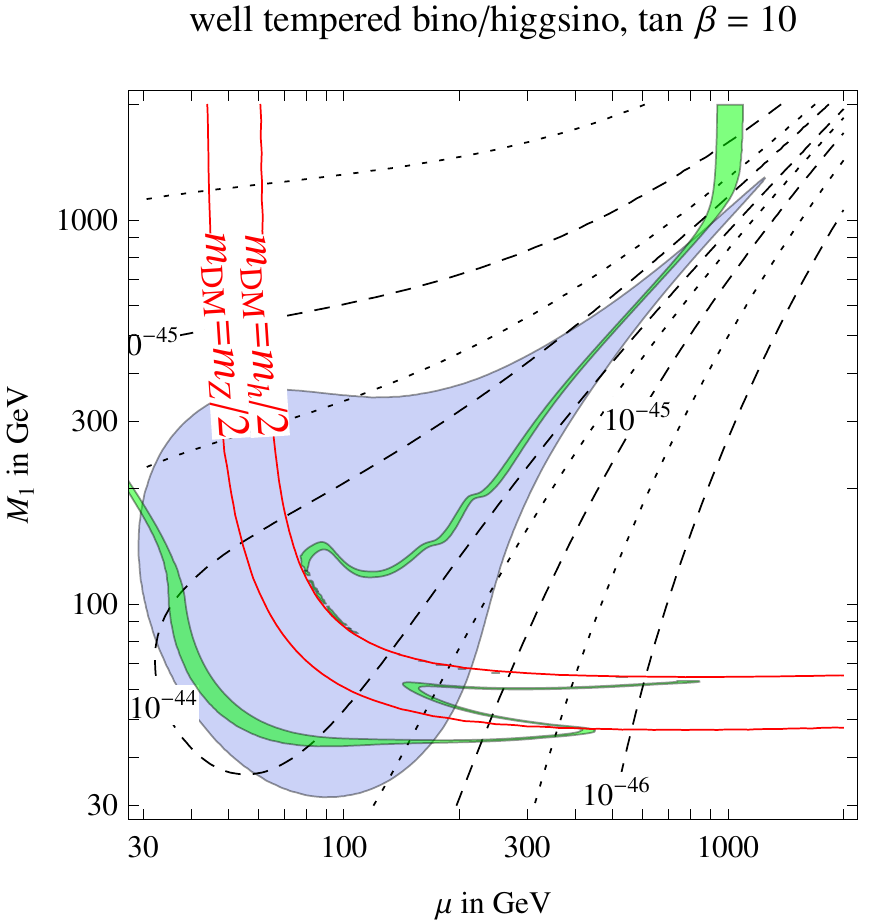}$$
\caption{\em (Update of fig.\fig{WT}).
Model independent ``well-tempered" neutralino scenario for $m_h =125\GeV$.
The $3\sigma$ range for the cosmological DM abundance is reproduced within the
green strip.  The gray region is excluded by {\sc Xenon}100~\cite{Xenon100}.
\label{fig:WT2012}}
%\end{figure}
%\begin{figure}[p]
$$%\includegraphics[width=0.45\textwidth]{CMSSMDMsigma}\qquad
\includegraphics[width=0.45\textwidth]{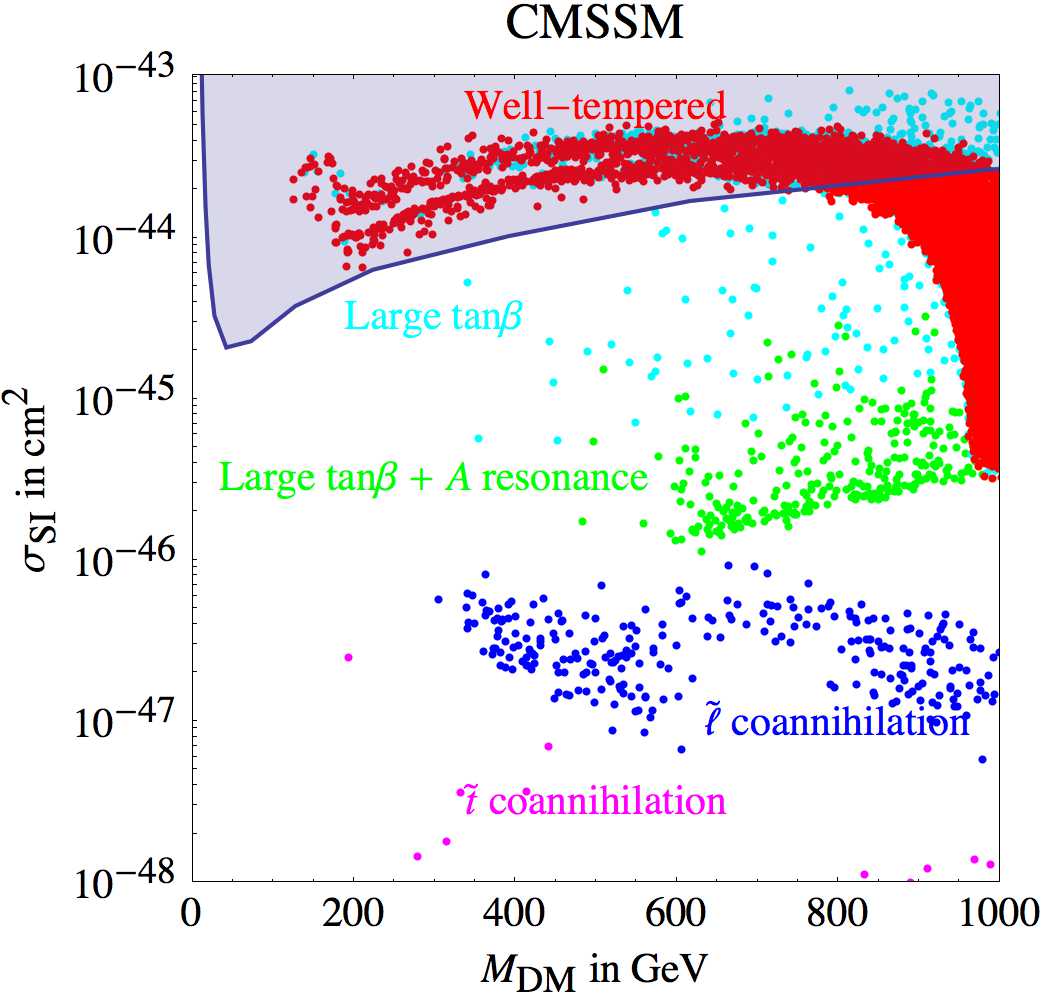}$$
\caption{\em (Update of fig.\fig{CMSSMfits2}).
{The ($M_{\rm DM},\sigma_{\rm SI}$) plane in the CMSSM}.
%In the left panel we show the global fit:
%the yellow regions surrounded by continuous contours are the best fit including the {\sc Xenon}100 and LHC data, at
%$68,95,99.7\%$ confidence levels for 2 d.o.f.  The red (blue) regions surrounded by dashed contours are the corresponding regions now excluded by {\sc Xenon}100 (LHC).
Points with $\Delta\chi^2<5^2$, colored according to the DM annihilation mechanism.
%The red dots in the upper region excluded by the {\sc Xenon}100 correspond to the ``well-tempered" neutralino, green via the heavy Higgs resonance, cyan via neutral Higgses with $\tan\beta$-enhanced couplings,
%blue via slepton co-annihilations, magenta via stop co-annihilations.
\label{fig:CMSSMfits2012}}
\end{figure}

Coming to supersymmetry,
fig.\fig{WT2012} shows that the well-tempered bino/higgsino considered in section~\ref{welltempered}
has been excluded.  We recomputed the DM detect detection cross section $\sigma_{\rm SI}$ now
assuming the measured Higgs mass, and we find that the `well-tempered' region along $M_1\approx |\mu|$
(green strip in the figure, where the DM thermal relic density matches the observed cosmological DM density)
is excluded.   The allowed region at $|\mu|\approx$ 1 TeV corresponds to the pure-Higgsino limit.
The allowed region at $M_{\rm DM}\approx m_h/2$ corresponds to the Higgs resonance region;
this possibility is however excluded, in the context of models with gaugino unification, because it implies a too light gluino,
excluded by LHC bounds.

Fig.\fig{CMSSMfits2012} shows in the ($M_{\rm DM},\sigma_{\rm SI}$) plane the CMSSM points that lead to the observed
DM abundance and are compatible with present data, including the recent Higgs mass measurement. 
From a theoretical point of view the CMSSM is a sort of `spherical cow'; most of its parameter space has now been excluded, and we are 
here focusing on its remaining peculiar tails and pieces.
In particular, the DM abundance can be explained by alternative fine-tuned mechanisms,
here plotted as different colours.  
In decreasing order of $\sigma_{\rm SI}$,
red dots correspond to the `well-tempered' bino/higgsino (excluded), extending into the pure-Higgsino limit.
Cyan dots correspond to DM annihilations via couplings enhanced by $\tan\beta\approx 50$.
Green dots correspond to annihilations via heavy Higgs resonances with large $\tan\beta$. 
Blue points correspond to slepton co-annihilation and magenta points to stop co-annihilation.
No real best-fit region emerges.


\footnotesize
\begin{multicols}{2}
 
\end{multicols}
\label{10out}

\end{document}